\pdfoutput=1
\documentclass[12pt]{article}
\usepackage{amsmath}
\usepackage{amssymb,bm}
\usepackage{graphicx,psfrag,epsf,subcaption}
\usepackage{enumerate}
\usepackage{natbib}
\usepackage{url} 
\usepackage{multirow}
\newcommand{\blind}{0}

\def\bfbeta{\bm{\beta}}
\def\bfeta{\bm{\eta}}
\def\bfgamma{\bm{\gamma}}

\begin{document}

\def\spacingset#1{\renewcommand{\baselinestretch}%
{#1}\small\normalsize} \spacingset{1}


\if0\blind
{
  \title{\bf BIVAS: A scalable Bayesian method for bi-level variable selection with applications}
  \author{Mingxuan Cai\\
    Department of Mathematics, Hong Kong Baptist University\\
    Mingwei Dai \\
    School of Mathematics and Statistics, Xi'an Jiaotong University\\
    Jingsi Ming\\
    Department of Mathematics, Hong Kong Baptist University\\
    Heng Peng\\
    Department of Mathematics, Hong Kong Baptist University\\
    Jin Liu\\
    Centre for Quantitative Medicine, Duke-NUS Medical School\\
    Can Yang\\
    Department of Mathematics,\\
    The Hong Kong University of Science
and Technology\\
    }
  \maketitle
} \fi

\if1\blind
{
  \bigskip
  \bigskip
  \bigskip
  \begin{center}
    {\LARGE\bf Title}
\end{center}
  \medskip
} \fi

\bigskip
\begin{abstract}
In this paper, we consider a Bayesian bi-level variable selection problem in high-dimensional regressions. In many practical situations, it is natural to assign group membership to each predictor. Examples include that genetic variants can be grouped at the gene level and a covariate from different tasks naturally forms a group. Thus, it is of interest to select important groups as well as important members from those groups. The existing Markov Chain Monte Carlo (MCMC) methods are often computationally intensive and not scalable to large data sets. To address this problem, we consider variational inference for bi-level variable selection (BIVAS). In contrast to the commonly used mean-field approximation, we propose a hierarchical factorization to approximate the posterior distribution, by utilizing the structure of bi-level variable selection. Moreover, we develop a computationally efficient and fully parallelizable algorithm based on this variational approximation. We further extend the developed method to model data sets from multi-task learning. The comprehensive numerical results from both simulation studies and real data analysis demonstrate the advantages of BIVAS for variable selection, parameter estimation and computational efficiency over existing methods. The method is implemented in R package `bivas' available at https://github.com/mxcai/bivas.
\end{abstract}

\noindent%
\vfill

\newpage
\spacingset{1.45} 
\section{Introduction}
\label{sec:intro}

Variable selection plays an important role in modern data analysis with the ever-increasing number of variables, where it is often assumed that only a small proportion of variables are relevant to the response variable \citep{hastie2015statistical}. In many real applications, this sparse pattern could be more complicated. In this paper, we consider a class of regression problems in which the grouping structure of the variables naturally exists. Examples include, but not limited to, the categorical predictors that are often represented by a group of indicators and continuous predictors that can be expressed by a group of basis functions. We assume that only a proportion of groups are relevant to the response variable and within each relevant group, only a subset of variables is relevant. Hence we consider a bi-level variable selection problem, i.e., variable selection at both the individual and group levels \citep{breheny2009penalized}.

There have been rich literatures on variable selection \citep{tibshirani1996,fan2001variable,zhang2010nearly,yuan2006model}, but majority of them focus on variable selection at the individual level, including penalized merhods, such as Lasso \citep{tibshirani1996}, SCAD \citep{fan2001variable} and MCP \citep{zhang2010nearly}, and Bayesian vairable selection methods based on sparsity-promoting priors, such as Laplace-like priors \citep{figueiredo2003adaptive,bae2004gene,yuan2005efficient,park2008bayesian} and spike-slab priors \citep{mitchell1988bayesian,george1993variable,madigan1994model,george1997approaches}. To perform variable selection at the group level, the group Lasso \citep{yuan2006model} introduced the $L_1$-$L_2$ norm penalty to group variables and perform group selection using the $L_1$ norm. CAP \citep{zhao2009composite} generalized this idea to be the $L_1$-$L_{\gamma}$ norm, where $\gamma\in[1,+\infty)$. Under the Bayesian framework, this is achieved by specifying the prior over a whole group of variables \citep{kyung2010penalized,raman2009bayesian,xu2015bayesian}.

The group variable selection methods usually act like Lasso at the group level and variables are selected in the `all-in or all-out' manner. However, these methods does not yield sparsity within a group, i.e. if a group is selected, all variables within that group will be non-zero. To conduct variable selection at both the individual and group levels, various methods have been proposed for bi-level selection from different perspectives including both penalized and Bayesian methods. Penalized methods often consider a composition of two penalties. The group bridge \citep{huang2009group} adopts a bridge penalty on the group level and the $L_{1}$ penalty on the variable level. Hierarchical Lasso \citep{zhou2010group} can be viewed as a special case of group bridge with bridge index fixed at $0.5$. Under certain regularity conditions, the global group bridge solution is proved to be group selection consistent. However, the singularity nature of these penalties at $0$ potentially complicates the optimization in practice. The composite MCP (cMCP) \citep{breheny2009penalized} and group exponential Lasso (GEL) \citep{breheny2015group} proposed to apply their penalty at both levels in a manner that puts less penalization as the absolute value of a coefficient becomes larger. 
On the other hand, Bayesian methods usually assume a spike-slab prior on variables at both the individual and group levels to promote bi-level sparsity. Despite the convenience of using Bayesian methods to depict hierarchical struture among variables, such posteriors are usually intractable. Hence, current literatures mainly rely on sampling methods to approximate the posterior distribution, such as Markov Chain Monte Carlo (MCMC) \citep{xu2015bayesian,chen2016bayesian}. The computational costs of these methods become very expensive in the presence of a large number of variables.

In this paper, we propose a scalable Bayesian method for \underline{bi}-level \underline{va}riable \underline{s}election (BIVAS). Instead of using MCMC, we adopt variational inference, which greatly reduces computational cost and makes our algorithm scalable. In contrast to standard mean-field variational approximation, we propose a hierarchically factorizable approximation, making use of the special structure of bi-level variable selection. A computationally efficient variational expectation-maximiztion (EM) algorithm is developed to handle large data sets. Moreover, we extend our approach to handle a class of multi-task learning. We further use comprehensive simulation studies to demonstrate that BIVAS can significantly outperform its alternatives in term of variable selection, prediction accuracy and computational efficiency.

The remainder of this article is organized as follows. In Section 2, we describe both model settings and algorithms. In particular, we show the rationale to improve the computational efficiency. We further discuss the way of extending our approach to multi-task learning. In Section 3, we evaluated the performance of BIVAS based on comprehensive simulation studies, especially checked the cases variational assumptions are violated. The experimental results show that BIVAS can stably outperform its alternatives in various settings. Then we applied BIVAS to three real data examples. We conclude the paper with a short discussion in Section 4.

\section{Methods}
\label{sec:meth}

\subsection{Regression with BIVAS}

\subsubsection{Model Setting}
Suppose we have collected dataset $\{\mathbf{y}, \mathbf{Z}, \mathbf{X}\}$ with sample size $n$, where $\mathbf{y}\in\mathbb{R}^n$ is the vector of response variable, $\mathbf{Z}\in\mathbb{R}^{n\times r}$ is the design matrix of $r$ columns including an intercept and a few covariates ($r<n$) and $\mathbf{X}\in\mathbb{R}^{n\times p}$ is the design matrix of $p$ predictors. Besides, each of the $p$ variables in $\mathbf{X}$ is labeled with one of $K$ known groups, where the number of variables in group $k$ is denoted by $l_k$ and $\sum_{k=1}^{K}l_{k}=p$. We consider the following linear model that links $\mathbf{y}$ to $\mathbf{Z}$ and $\mathbf{X}$:
\begin{equation}
  \mathbf{y}=\mathbf{Z}\bm{\omega}+\mathbf{X}\bm{\beta}+\mathbf{e},
\end{equation}
where $\bm{\omega}\in\mathbb{R}^{r}$ is a vector of fixed effects, $\bm{\beta}\in\mathbb{R}^{p}$ is a vector of random effects, and $\mathbf{e}\in\mathbb{R}^{n}$ is a vector of independent noise. We assume $\mathbf{e}\sim \mathcal{N}(\mathbf{0},\sigma_{e}^{2}\mathbf{I}_n)$, where $\mathbf{I}_n$ is the $n$-by-$n$ identity matrix. Under this model, the bi-level selection aims to identify non-zero entries of $\bm{\beta}$ at both the group and individual-variable levels. For this reason, we introduce two binary variables: $\eta_{k}$ indicates whether group $k$ is active ($\eta_k=1$) or not ($\eta_k=0$); and $\gamma_{jk}$ indicates whether the $j$-th variable in group $k$ is zero ($\gamma_{jk}=0$) or not ($\gamma_{jk}=1$). Hence, we introduce a bi-level spike-slab prior on $\bm{\beta}$:
\begin{equation}
\beta_{jk}|\eta_{k},\gamma_{jk};\sigma_{\beta}^{2}\sim  \begin{cases}
                \mathcal{N}(\beta_{jk}|0,\sigma_{\beta}^{2}) &\text{if}\ \eta_{k}=1,\ \gamma_{jk}=1,\\
                \delta_{0}(\beta_{jk}) &\text{otherwise} ,
                \end{cases}
\end{equation}
where $\mathcal{N}(\beta_{jk}|0,\sigma_{\beta}^{2})$ denotes the Gaussian distribution with mean $0$ and variance $\sigma_{\beta}^{2}$ and $\delta_{0}(\beta_{jk})$ denotes a Dirac function at zero. This bi-level structure means that $\beta_{jk}$ is drawn from $\mathcal{N}(0,\sigma_{\beta}^{2})$ if and only if both the $k$-th group and its $j$-th variable are included in the model. Let Pr$(\eta_{k}=1)=\pi$ and Pr$(\gamma_{jk}=1)=\alpha$ be the prior inclusion probability of groups and variables, respectively.

The presence of Dirac function may introduce additional troubles in algorithm derivation. To get rid of the Dirac function, we re-parameterize the model as following:
\begin{eqnarray}
  \beta_{jk}|\sigma_{\beta}^{2} \sim \mathcal{N}(0,\sigma_{\beta}^{2}),\ \ 
  \gamma_{jk}|\alpha \sim \alpha^{\gamma_{jk}}(1-\alpha)^{1-\gamma_{jk}} ,\ \ 
  \eta_{k}|\pi \sim \pi^{\eta_{k}}(1-\pi)^{1-\eta_{k}}.
\end{eqnarray}
Consequently, the prior of $\beta_{jk}$ does not depend on $\gamma_{jk}$ and $\eta_{k}$ any more, and the product $\eta_{k}\gamma_{jk}\beta_{jk}$ form a new random variable exactly distributed as given in (2). We shall use the re-parameterized version through the paper.

Let $\bm{\theta}=\{\alpha,\pi,\sigma_{\beta}^{2},\sigma_{e}^{2},\bm{\omega}\}$ be the collection of model parameters and $\{\bm{\beta}, \bm{\gamma}, \bm{\eta}\}$ be the set of latent variables. The joint probabilistic model is
\begin{eqnarray}
  \begin{aligned}
  &\mathrm{Pr}(\mathbf{y},\bm{\eta},\bm{\gamma},\bm{\beta}|\mathbf{X},\mathbf{Z};\bm{\theta}) 
  =\mathrm{Pr}(\mathbf{y}|\bm{\eta},\bm{\gamma},\bm{\beta},\mathbf{X},\mathbf{Z},\bm{\theta})\mathrm{Pr}(\bm{\eta},\bm{\gamma},\bm{\beta}|\bm{\theta})  \\
   =&\mathcal{N}(\mathbf{y}|\mathbf{Z}\bm{\omega}+\sum_{k}^{K}\sum_{j}^{l_{k}}\eta_{k}\gamma_{jk}\beta_{jk}\mathbf{x}_{jk},\sigma_e^2)\prod_{k=1}^{K}\pi^{\eta_{k}}(1-\pi)^{1-\eta_{k}}\prod_{j=1}^{l_{k}}\mathcal{N}(0,\sigma_{\beta}^{2})\alpha^{\gamma_{jk}}(1-\alpha)^{1-\gamma_{jk}},
  \end{aligned}
\end{eqnarray}
where $\mathbf{x}_{jk}$ is a column of $\mathbf{X}$ corresponding to the $j$-th variable in the $k$-th group. The goal is to obtain the estimate of $\bm{\theta}$, $\hat{\bm{\theta}}$, by optimizing the marginal likelihood 
\begin{eqnarray}
  \log\ \mathrm{Pr}(\mathbf{y}|\mathbf{X},\mathbf{Z};\bm{\theta})=\log\sum_{\bm{\gamma}}\sum_{\bm{\eta}}\int_{\bm{\beta}}\mathrm{Pr}(\mathbf{y},\bm{\eta},\bm{\gamma},\bm{\beta}|\mathbf{X},\mathbf{Z};\bm{\theta})d\bm{\beta},
\end{eqnarray}
and evaluate the posterior 
\begin{eqnarray}
\mathrm{Pr}(\bm{\eta},\bm{\gamma},\bm{\beta}|\mathbf{y},\mathbf{X},\mathbf{Z};\hat{\bm{\theta}})=\frac{\mathrm{Pr}(\mathbf{y},\bm{\eta},\bm{\gamma},\bm{\beta}|\mathbf{X},\mathbf{Z};\hat{\bm{\theta}})}{\mathrm{Pr}(\mathbf{y}|\mathbf{X},\mathbf{Z};\hat{\bm{\theta}})}.
\end{eqnarray}

\subsubsection{Algorithm}
Conventionally, the model involving latent variables is often solved by the Expectation-Maximization (EM) algorithm. However, the standard EM algorithm cannot be applied here due to the difficulty of the E-step caused by the combinatorial nature of $\bm{\gamma}$ and $\bm{\eta}$. Alternatively, we propose a variational EM algorithm via approximate Bayesian inference \citep{bishop2006pattern}.

To apply variational approximation, we first define $q(\bm{\eta},\bm{\gamma},\bm{\beta})$ as an approximated distribution of posterior $\mathrm{Pr}(\bm{\eta},\bm{\gamma},\bm{\beta}|\mathbf{y},\mathbf{X},\mathbf{Z};\bm{\theta})$. Then we can obtain the lower bound of log-marginal likelihood by Jensen's inequality:
\begin{eqnarray}
  \begin{aligned}
  \log\ p(\mathbf{y}|\mathbf{X},\mathbf{Z};\bm{\theta})&=\log\sum_{\bm{\gamma}}\sum_{\bm{\eta}}\int_{\bm{\beta}}\mathrm{Pr}(\mathbf{y},\bm{\eta},\bm{\gamma},\bm{\beta}|\mathbf{X},\mathbf{Z};\bm{\theta})d\bm{\beta}\\
  &\geq\sum_{\bm{\gamma}}\sum_{\bm{\eta}}\int_{\bm{\beta}}q(\bm{\eta},\bm{\gamma},\bm{\beta})\log \frac{\mathrm{Pr}(\mathbf{y},\bm{\eta},\bm{\gamma},\bm{\beta}|\mathbf{X},\mathbf{Z};\bm{\theta})}{q(\bm{\eta},\bm{\gamma},\bm{\beta})}d\bm{\beta}\\
  &=\mathbb{E}_q[\log\mathrm{Pr}(\mathbf{y},\bm{\eta},\bm{\gamma},\bm{\beta}|\mathbf{X},\mathbf{Z};\bm{\theta})-\log q(\bm{\eta},\bm{\gamma},\bm{\beta})]\\
  &\equiv \mathcal{L}(q),
  \end{aligned}
\end{eqnarray}
where the equality holds if and only if $q(\bm{\eta},\bm{\gamma},\bm{\beta})=\mathrm{Pr}(\bm{\eta},\bm{\gamma},\bm{\beta}|\mathbf{y},\mathbf{X},\mathbf{Z};\bm{\theta})$. Then, we can iteratively maximize $\mathcal{L}(q)$ instead of working with the marginal likelihood directly. Conventionally, $q$ is often assumed to be fully factorizable based on the mean-field theory \citep{bishop2006pattern}. As there is hierarchical structure between the group level and the variable level, here we propose a novel variational distribition to accommodate the bi-level variable selection. Specifically, we consider the the following hierarchically structured distribution as an approximation to posterior $\mathrm{Pr}(\bm{\eta},\bm{\gamma},\bm{\beta}|\mathbf{y},\mathbf{X},\mathbf{Z})$:
\begin{equation}
q(\bm{\eta},\bm{\gamma},\bm{\beta})=\prod_{k}^{K}\left(q(\eta_{k})\prod_{j}^{l_{k}}(q(\beta_{jk}|\eta_{k},\gamma_{jk})q(\gamma_{jk}))\right).
\end{equation}

Without any other assumptions, we can show (with details in Supplementary) that  the optimal solution of $q$ is given as:
\begin{equation}
q(\bm{\eta},\bm{\gamma},\bm{\beta})=\prod_{k}^{K}\left(\pi_{k}^{\eta_{k}}(1-\pi_{k})^{1-\eta_{k}}\prod_{j}^{l_{k}}\left(\alpha_{jk}^{\gamma_{jk}}(1-\alpha_{jk})^{1-\gamma_{jk}}\mathcal{N}(\mu_{jk},s_{jk}^{2})^{\eta_{k}\gamma_{jk}}\mathcal{N}(0,\sigma_{\beta}^2)^{1-\eta_{k}\gamma_{jk}}\right)\right),
\end{equation}
where
\begin{eqnarray}
  \begin{aligned}
  s_{jk}^{2}&=\frac{\sigma_{e}^{2}}{\mathbf{x}_{jk}^{T}\mathbf{x}_{jk}+\frac{\sigma_{e}^{2}}{\sigma_{\beta}^{2}}},\\
  \mu_{jk}&=\frac{\mathbf{x}_{jk}^{T}(\mathbf{y}-\mathbf{Z}\bm{\omega})-\sum_{k^{\prime}\neq k}^{K}\sum_{j}^{l_{k}}\mathbb{E}_{jk^{\prime}}[\eta_{k^{\prime}}\gamma_{jk^{\prime}}\beta_{jk^{\prime}}]\mathbf{x}_{jk^{\prime}}^{T}\mathbf{x}_{jk}-\sum_{j^{\prime}\neq j}^{l_{k}}\mathbb{E}[\gamma_{j^{\prime}k}\beta_{j^{\prime}k}]\mathbf{x}_{j^{\prime}k}^{T}\mathbf{x}_{jk}}{\mathbf{x}_{jk}^{T}\mathbf{x}_{jk}+\frac{\sigma_{e}^{2}}{\sigma_{\beta}^{2}}},
  \end{aligned}
\end{eqnarray}

\begin{eqnarray}
  \begin{aligned}
  \pi_{k}=\frac{1}{1+\mathrm{exp}(-u_{k})},\ 
  \mathrm{with}\ u_{k}=\log\frac{\pi}{1-\pi}+\frac{1}{2}\sum_{j}^{l_{k}}\alpha_{jk}\left(\log\frac{s_{jk}^{2}}{\sigma_{\beta}^{2}}+\frac{\mu_{jk}^{2}}{s_{jk}^{2}}\right),
  \end{aligned}\\
  \begin{aligned}
  \alpha_{jk}=\frac{1}{1+\mathrm{exp}(-v_{jk})},\ 
  \mathrm{with}\  v_{jk}=\log\frac{\alpha}{1-\alpha}+\frac{1}{2}\pi_{k}\left(\log\frac{s_{jk}^{2}}{\sigma_{\beta}^{2}}+\frac{\mu_{jk}^{2}}{s_{jk}^{2}}\right).
  \end{aligned}
\end{eqnarray}

By inspections of Equations (8) and (9), we have $q(\eta_{k}=1)=\pi_{k}$ and $q(\gamma_{jk}=1)=\alpha_{jk}$, which can be viewed as approximations to the posterior distibutions $\Pr(\eta_k=1|\mathbf{y},\mathbf{X},\mathbf{Z};\bm{\theta})$ and $\Pr(\gamma_{jk}=1|\mathbf{y},\mathbf{X},\mathbf{Z};\bm{\theta})$, repectively. Similarly, $q(\beta_{jk}|\eta_{k}\gamma_{jk}=1)=\mathcal{N}(\mu_{jk},s_{jk}^{2})$ can be interpreted as the variational approximation to $\Pr(\beta_{jk}|\eta_{k}\gamma_{jk}=1,\mathbf{y},\mathbf{X},\mathbf{Z};\bm{\theta})$, which is the conditional posterior distribution of $\beta_{jk}$ given it is selected in both the group level and the variable level. Accordingly, $q(\beta_{jk}|\eta_{k}\gamma_{jk}=0)=\mathcal{N}(0,\sigma_{\beta}^{2})$ approximates $\Pr(\beta_{jk}|\eta_{k}\gamma_{jk}=0,\mathbf{y},\mathbf{X},\mathbf{Z};\bm{\theta})$, corresponding to the case when $\beta_{jk}$ is irrelevant in either of the two levels.

Note that the form of variational parameters provides an intuitive interpretation. Group-level posterior inclusion probability $\pi_{k}$ and variable-level posterior inclusion probability $\alpha_{jk}$ can be viewed as their prior inclusion probability ($\pi$, $\alpha$) updated by data-driven information. Furthermore, $\pi_k$ and $\alpha_{jk}$ are interdependent. On one hand, if more and more $\alpha_{jk}$ within the $k$-th group become closer to one, then $\pi_k$ will be closer to one, as seen in Equation (11). On the other hand, if $\pi_k$ increases, then the variables in the $k$-th group are more likely to be selected, see Equation (12).

With Equation (9), the lower bound $\mathcal{L}(q)$ can be evaluated analytically. By setting the derivative of $\mathcal{L}(q)$ with respect to $\bm{\theta}$ to be zero, we have the updating equations for parameter estimation:

  \begin{eqnarray}
 \begin{aligned}
  \sigma_{e}^{2}=&\frac{||\mathbf{y}-\mathbf{Z}\bm{\omega}-\sum_{k}^{K}\sum_{j}^{l_{k}}\pi_{k}\alpha_{jk}\mu_{jk}\mathbf{x}_{jk}||^{2}}{n}\\
&+\frac{\sum_{k}^{K}\sum_{j}^{l_{k}}[\pi_{k}\alpha_{jk}(s_{jk}^{2}+\mu_{jk}^{2})-(\pi_{k}\alpha_{jk}\mu_{jk})^{2}]\mathbf{x}_{jk}^{T}\mathbf{x}_{jk}}{n}\\
&+\frac{\sum_{k}^{K}(\pi_{k}-\pi_{k}^{2})[\sum_{j}^{l_{k}}\sum_{j^{\prime}}^{l_{k}}\alpha_{j^{\prime}k}\mu_{j^{\prime}k}\alpha_{jk}\mu_{jk}]\mathbf{x}_{j^{\prime}k}^{T}\mathbf{x}_{jk}}{n},\\
 \sigma_{\beta}^{2}=&\frac{\sum_{k}^{K}\sum_{j}^{l_{k}}\pi_{k}\alpha_{jk}(s_{jk}^{2}+\mu_{jk}^{2})}{\sum_{k}^{K}\sum_{j}^{l_{k}}\pi_{k}\alpha_{jk}},\\
 \alpha=&\frac{1}{p}\sum_{k}^{K}\sum_{j}^{l_{k}}\alpha_{jk},\\
 \pi=&\frac{1}{K}\sum_{k}^{K}\pi_{k},\\
 \bm{\omega}=&(\mathbf{Z}^{T}\mathbf{Z})^{-1}\mathbf{Z}^{T}(\mathbf{y}-\sum_{k}^{K}\sum_{j}^{l_{k}}\pi_{k}\alpha_{jk}\mu_{jk}\mathbf{x}_{jk}).\\
 \end{aligned}
 \end{eqnarray}

To summarize, the algorithm can be regarded as a variational extension of the EM algorithm. At E-step, the lower bound $\mathcal{L}(q)$ is obtained by evaluating the expectation w.r.t variational posterior $q$. At M-step, the current $\mathcal{L}(q)$ is optimized w.r.t model parameters in $\bm{\theta}$. As a result, the lower bound increases at each iteration and the convergence is guaranteed.

\subsection{Multi-task learning with BIVAS}

\subsubsection{Model Setting}
In this section, we consider bi-level variable selection in multi-task learning. In real applications, some related regression tasks may have similar patterns in the effects of predictor variables. A joint model that analyze all such related tasks simultaneously can efficiently increase statistical power, which is called multi-task learning \citep{caruana1998multitask}. As we shall see later, a class of multi-task regression problem can be naturally solved by BIVAS with proper adjustment for the likelihood. To avoid ambiguity, we refer to the model described in Section 2.1 as `group BIVAS' and the one discussed in this section as `multi-task BIVAS'.

Suppose we have collected dataset $\{\mathbf{y}, \mathbf{Z}, \mathbf{X}\}=\{\mathbf{y}_j, \mathbf{Z}_j, \mathbf{X}_j\}_{j=1}^L$ from $L$ related regression tasks, each of which has sample size $n_j$. In practice, $\mathbf{y}_j\in\mathbb{R}^{n_j}$ is the the reponse vector of $j$-th task from $n_j$ individuals; $\mathbf{Z}_{j}\in\mathbb{R}^{n_{j}\times r}$ includes an intercept and a few shared covariates; $\mathbf{X}_{j}\in\mathbb{R}^{n_{j}\times K}$ is the design matrix of $K$ shared predictors. We relate $\mathbf{y}_j$ to $\mathbf{X}_j$ and $\mathbf{Z}_j$ using the following linear mixed model:
\begin{equation}
  \mathbf{y}_{j}=\mathbf{Z}_{j}\bm{\omega}_{j}+\mathbf{X}_{j}\bm{\beta}_{j}+\mathbf{e}_{j},\ j=1,\dots,L,
\end{equation}
where $\bm{\omega}_{j}\in\mathbb{R}^{r}$ is the vector of fixed effects, $\bm{\beta}_{j}\in\mathbb{R}^{K}$ is the vector of random effects and $\mathbf{e}_{j}\in\mathbb{R}^{n_{j}}$ is the vector of independent noise with $\mathbf{e}_{j}\sim \mathcal{N}(\mathbf{0},\sigma_{e_{j}}^{2}\mathbf{I}_{n_j})$. For convenience, we denote $\bm{\beta}=[\bm{\beta}_1,...,\bm{\beta}_L]\in\mathbb{R}^{K\times L}$ and $\beta_{jk}$ be $k$-th entry in $\bm{\beta}_{j}$. Clearly, it is not reasonable to assume that all shared predictors are relevant to all reponses, especially when $K$ is large. A more reasonable assumpotion is that the majority of predictors are irrelevant to all the responses and only a few of them are relevant with many responses. With this assumption, it is natural to treat each shared predictor as a group across different task $l$, which corresponds to a row of $\bm{\beta}$. Then the group-level selection aims at excluding variables which are irrelevant to all responses and the individual-level selection further identifies fine-grained relevance between variables and response of specific task. For this purpose, we introduce two binary variables: $\eta_k$ indicates whether the $k$-th row of $\bm{\beta}$ is active or not and $\gamma_{jk}$ indicates whether $\bm{\beta}_{jk}$ is zero or not. Then the bi-level spike-slab prior on $\bm{\beta}$ is introduced by:
\begin{equation}
\beta_{jk}|\eta_k,\gamma_{jk};\sigma_{\beta_j}^2\sim  \begin{cases}
                \mathcal{N}(\beta_{jk}|0,\sigma_{\beta_{j}}^{2}) &\text{if}\ \eta_{k}=1,\ \gamma_{jk}=1,\\
                \delta_{0}(\beta_{jk}) &\text{otherwise},
                \end{cases}
\end{equation}
where prior inclusion probabilities are defined as Pr$(\eta_{k}=1)=\pi$ and Pr$(\gamma_{jk}=1)=\alpha$.

Again we re-parameterize the model to remove the Dirac function:
\begin{eqnarray}
  \beta_{jk}|\sigma^{2}_{\beta_{j}} \sim \mathcal{N}(0,\sigma_{\beta_{j}}^{2}), \ \
  \gamma_{jk}|\alpha \sim \alpha^{\gamma_{jk}}(1-\alpha)^{1-\gamma_{jk}},\ \ 
  \eta_{k}|\pi \sim \pi^{\eta_{k}}(1-\pi)^{1-\eta_{k}}.
  \end{eqnarray}

Let $\bm{\theta}=\{\alpha,\pi,\sigma_{\beta_{j}}^{2},\sigma_{e_{j}}^{2},\bm{\omega}_{j}\}_{j=1}^{L}$ be the collection of parameters under the multi-task model. The joint probabilistic model is
\begin{eqnarray}
  \begin{aligned}
  &\mathrm{Pr}(\mathbf{y},\bm{\eta},\bm{\gamma},\bm{\beta}|\mathbf{X},\mathbf{Z};\bm{\theta}) 
  =\mathrm{Pr}(\mathbf{y}|\bm{\eta},\bm{\gamma},\bm{\beta},\mathbf{X},\mathbf{Z},\bm{\theta})\mathrm{Pr}(\bm{\eta},\bm{\gamma},\bm{\beta}|\bm{\theta})  \\
   =&\prod_{j=1}^{L}\mathcal{N}(\mathbf{y}_{j}|\mathbf{Z}_{j}\bm{\omega}_{j}+\sum_{k}^{K}\eta_{k}\gamma_{jk}\beta_{jk}\mathbf{x}_{jk},\sigma_{e_j}^2)\prod_{k=1}^{K}\pi^{\eta_{k}}(1-\pi)^{1-\eta_{k}}\prod_{j=1}^{L}\mathcal{N}(0,\sigma_{\beta_{j}}^{2})\alpha^{\gamma_{jk}}(1-\alpha)^{1-\gamma_{jk}},
  \end{aligned}
\end{eqnarray} 
where $\mathbf{x}_{jk}$ is the $k$-th column of $\mathbf{X}_{j}$, corresponding to the $k$-th variabe in the $j$-th task. Our goal is to maximize the marginal likelihood, which is of the same form as Equation (5), and evaluate the posterior distribution of $\beta_{jk}$.

\subsubsection{Algorithm}
The variational EM algorithm of multi-task BIVAS is straightforward following the similar procedure in 2.1.2. We leave the details in the supplementary document. In summary, we have
 \begin{eqnarray}
   \begin{aligned}
    s_{jk}^{2}&=\frac{\sigma_{e_{j}}^{2}}{\mathbf{x}_{jk}^{T}\mathbf{x}_{jk}+\frac{\sigma_{e_{j}}^{2}}{\sigma_{\beta_{j}}^{2}}}\\
   \mu_{jk}&=\frac{\mathbf{x}_{jk}^{T}(\mathbf{y}_{j}-\mathbf{Z}_{j}\bm{\omega}_{j}-\tilde{\mathbf{y}}_{jk})}{\mathbf{x}_{jk}^{T}\mathbf{x}_{jk}+\frac{\sigma_{e_{j}}^{2}}{\sigma_{\beta_{j}}^{2}}}\\
   \pi_{k}&=\frac{1}{1+\mathrm{exp}(-u_{k})},\ \mathrm{where}\  u_{k}=\mathrm{log}\frac{\pi}{1-\pi}+\frac{1}{2}\sum_{j}^{L}\alpha_{jk}\left(\mathrm{log}\frac{s_{jk}^{2}}{\sigma_{\beta_{j}}^{2}}+\frac{\mu_{jk}^{2}}{s_{jk}^{2}}\right)\\
   \alpha_{jk}&=\frac{1}{1+\mathrm{exp}(-v_{k})},\ \mathrm{where}\  v_{k}=\mathrm{log}\frac{\alpha}{1-\alpha}+\frac{1}{2}\pi_{k}\left(\mathrm{log}\frac{s_{jk}^{2}}{\sigma_{\beta_{j}}^{2}}+\frac{\mu_{jk}^{2}}{s_{jk}^{2}}\right)\\
   \end{aligned}
 \end{eqnarray}
for E-step; and
  \begin{eqnarray}
 \begin{aligned}
  \sigma_{e_{j}}^{2}=&\frac{||\mathbf{y}_{j}-\mathbf{Z}_{j}\bm{\omega}_{j}-\sum_{k}^{K}\pi_{k}\alpha_{jk}\mu_{jk}\mathbf{x}_{jk}||^{2}}{N_{j}}\\
&+\frac{\sum_{k}^{K}[\pi_{k}\alpha_{jk}(s_{jk}^{2}+\mu_{jk}^{2})-(\pi_{k}\alpha_{jk}\mu_{jk})^{2}]\mathbf{x}_{jk}^{T}\mathbf{x}_{jk}}{N_{j}},\\
 \sigma_{\beta_{j}}^{2}=&\frac{\sum_{k}^{K}\pi_{k}\alpha_{jk}(s_{jk}^{2}+\mu_{jk}^{2})}{\sum_{k}^{K}\pi_{k}\alpha_{jk}},\\
 \alpha=&\frac{1}{p}\sum_{k}^{K}\sum_{j}^{L}\alpha_{jk},\\
 \pi=&\frac{1}{K}\sum_{k}^{K}\pi_{k},\\
 \bm{\omega}_{j}=&(\mathbf{Z}_{j}^{T}\mathbf{Z}_{j})^{-1}\mathbf{Z}_{j}^{T}(\mathbf{y}_{j}-\sum_{k}^{K}\pi_{k}\alpha_{jk}\mu_{jk}\mathbf{x}_{jk}),\\
 \end{aligned}
 \end{eqnarray}
 for M-step.

\subsection{Implementation Details}
After the convergence of algorithm, we can approximate the posterior inclusion probabilities by the variational approximation. For group BIVAS, the approximations are given by
 \begin{eqnarray*}
 \begin{aligned}
  \Pr(\eta_k=1|\mathbf{y},\mathbf{X},\mathbf{Z};\hat{\bm{\theta}})\approx q(\eta_k=1|\hat{\bm{\theta}})=\pi_k,\\
  \Pr(\gamma_{jk}=1|\mathbf{y},\mathbf{X},\mathbf{Z};\hat{\bm{\theta}})\approx q(\gamma_{jk}=1|\hat{\bm{\theta}})=\alpha_{jk}.\\ 
 \end{aligned}
 \end{eqnarray*}

These evaluations are based on parameter estimates $\hat{\bm{\theta}}$. However, as there is no guarantee of global optimal for the EM algorithm, the choice of initial value $\bm{\theta}^{0}$ is critical. A bad initial value will lead to a poor $\hat{\bm{\theta}}$. In our model, due to the existence of multiple latent variables ($\bm{\beta}$, $\bm{\eta}$, $\bm{\gamma}$), choosing a good initial value could be challenging. Here we consider the importance sampling suggested by varbvs \citep{carbonetto2012scalable}: we further introduce a prior over $\bm{\theta}$ and integrate over the value of $\bm{\theta}$ to obtain the final evaluations. In contrast to varbvs, we introduce prior only on the group sparsity parameter $\pi$. We first select $h$ values of $\pi$ ($\{\pi(i)\}_{i=1}^{h}$) such that $\log_{10}$ odds of $\pi(i)$ is uniformly distributed on $[-\log_{10}(K),0]$ which encourages group sparsity. With this additional setting, the new collection of parameters is then defined as $\bm{\theta}'=\{\bm{\theta}'_i\}_{i=1}^h$ with $\bm{\theta}'_i=\{\alpha,\pi(i),\sigma_\beta^2,\sigma_e^2,\bm{\omega}\}$; and the posterior inclusion probability can be approximated as follows:
 \begin{eqnarray}
 \begin{aligned}
  \Pr(\eta_k=1|\mathbf{y},\mathbf{X},\mathbf{Z}) \approx \int q(\eta_k=1|\bm{\theta}')\Pr(\bm{\theta}'|\mathbf{y},\mathbf{X},\mathbf{Z})d\bm{\theta}' \approx \frac{\sum_{i=1}^h q(\eta_k=1|\bm{\theta}_i')w(\bm{\theta}'_i)}{\sum_{i=1}^hw(\bm{\theta}'_i)},\\ 
  \Pr(\gamma_{jk}=1|\mathbf{y},\mathbf{X},\mathbf{Z}) \approx \int q(\gamma_{jk}=1|\bm{\theta}')\Pr(\bm{\theta}'|\mathbf{y},\mathbf{X},\mathbf{Z})d\bm{\theta}' \approx \frac{\sum_{i=1}^h q(\gamma_{jk}=1|\bm{\theta}_i')w(\bm{\theta}'_i)}{\sum_{i=1}^hw(\bm{\theta}'_i)},
 \end{aligned}
 \end{eqnarray}
where $w(\bm{\theta}_i')$ is the unnormalized importance weight for $i$-th component. For each of the two equations in (20), the first approximation is due to the variational inference; the second approximation is due to the importance sampling. Besides, $w(\bm{\theta}_i')$ can be approximated by exponential of $\mathcal{L}(q)$ given $\bm{\theta}'_i$ since $\mathcal{L}(q)$ takes similar shape to $\log\Pr(\mathbf{y}|\mathbf{X},\mathbf{Z};\bm{\theta})$ when the marginal likelihood is relatively large \citep{carbonetto2012scalable}. Hence, we can derive the final evaluation of posteriors:
 \begin{eqnarray}
 \begin{aligned}
  \Pr(\eta_k=1|\mathbf{y},\mathbf{X},\mathbf{Z}) \approx \sum_{i=1}^{h}\pi_{k}(i)\cdot \tilde{w}(i) \equiv \tilde{\pi}_k,\\
  \Pr(\gamma_{jk}=1|\mathbf{y},\mathbf{X},\mathbf{Z}) \approx \sum_{i=1}^{h}\alpha_{jk}(i)\cdot \tilde{w}(i) \equiv \tilde{\gamma}_{jk},\\
  \mathbb{E}(\beta_{jk}|\eta_k\gamma_{jk}=1,\mathbf{y},\mathbf{X},\mathbf{Z}) \approx \sum_{i=1}^{h}\mu_{jk}(i)\cdot \tilde{w}(i) \equiv \tilde{\mu}_{jk},\\
  \mathbb{E}(\eta_k\gamma_{jk}\beta_{jk}|\mathbf{y},\mathbf{X},\mathbf{Z})\approx \tilde{\pi}_k\tilde{\gamma}_{jk}\tilde{\mu}_{jk},
 \end{aligned}
 \end{eqnarray}
 where
\begin{equation*}
\tilde{w}(i)=\exp(w(\bm{\theta}'_i)-m),\ \ \ \ m=\max(w(\bm{\theta}'_i)).
\end{equation*}

Here we handle the normalization inside the exponantial so that the calculation is numerically stable. The same weighting evaluation applies to the parameters $\bm{\theta}'$. We can derive the same procedure for multi-task BIVAS as the one for group BIVAS. Although we need to run EM algorithm $h$ times in this procedure, each EM algorithm becomes more stable and converges in less iterations. In practice, $h=20\sim 40$ is often good enough for large scale data sets. Furthermore, taking the advantage of independence among $\pi(i)$'s, the $h$ procedures can be fully parallelized. Common solutions to parallelization are based on APIs such as OpenMP. These solutions, however, usually require the tasks to be allocated beforehand. In our model, this restriction may lead to inefficiency because the time of convergence for each procedure can be very different. Thus, we adopt a dynamic threading technique that can immediately allocates a new task to a thread once it has finished an old task. This technique greatly improves the efficiency of parallelization compared to OpenMP.

\subsection{Variable Selection and Prediction}
With the results obtained by importance sampling, we extract information from our model for the purpose of variable selection and prediction. Using the approximation of the posterior inclusion probability in (21), we can approximate local false discovery rate (fdr) of group $k$ by $fdr_{k}=1-\tilde{\pi}_{k}$ and fdr of $j$-th variable in $k$-th group by $fdr_{jk}=1-\tilde{\alpha}_{jk}$. Hence, by setting a commonly used threshold (e.g $fdr<0.05$), variables and groups with high posterior inclusion probability can be identified as relevant. Although the parameter estimates may not be accurate due to the variational approximation, the posterior means of latent variables appear to be accurate. We will verify this result later in the simulation.

In addition to variable selection, we can also predict $\hat{y}$ (or $\hat{y}_{j}$ for multi-task learning) with a new data $\{\mathbf{Z}^{new},\mathbf{X}^{new}\}$. Since $\mathbb{E}_q(\eta_{k}\gamma_{jk}\beta_{jk})\approx\tilde{\pi}_{k}\tilde{\alpha}_{jk}\tilde{\mu}_{jk}$ gives the estimate of effect size for the $jk$-th random effect, the predicted value is simply obtained by $\hat{y}=\sum_{r}\tilde{\omega}_r z_r^{new}+\sum_{k}\sum_{j}\tilde{\pi}_{k}\tilde{\alpha}_{jk}\tilde{\mu}_{jk}x^{new}_{jk}$ (in multi-task learning $\hat{y}_{j}=\sum_{r}\tilde{\omega}_{jr} z_{jr}^{new}+\sum_{k}\tilde{\pi}_{k}\tilde{\alpha}_{jk}\tilde{\mu}_{jk}x^{new}_{jk}$ for $j$-th task).

\section{Numerical Examples}
\label{sec:conc}

In this section, we gauged the performance of BIVAS in comparison with alternative methods using both simulation and real data analysis. In the spirit of reproducibility, all the simulation codes are made publicly availible at https://github.com/mxcai/sim-bivas.
\subsection{Simulation Study}
For group BIVAS, we compared it with varbvs \citep{carbonetto2012scalable}, cMCP \citep{breheny2009penalized}, and GEL \citep{breheny2015group}. The simulation data sets were generated as follows. The design matrix $\mathbf{X}$ was generated from normal distribution with autoregressive correlation $\rho^{|j-j^{\prime}|}$ between column $j$ and $j^{\prime}$. As the variational approximation assumes a hierarchically factorizable distribution, we selected $\rho\in\{-0.5,\ 0,\ 0.5\}$ to evaluate the influence of violation of this assumption. Next, we generated coefficients with different sparsity proportion at the group and individual levels: $(\pi,\alpha) \in \{(0.05,0.8),\ (0.1,0.4),\ (0.2,0.2),\ (0.4,0.1),\ (0.8,005)\}$. Note that the total sparsity was fixed at $\alpha\cdot\pi=0.04$ for different combinations of $\pi$ and $\alpha$. Finally, we controled the signal-to-noise ratio (SNR) at $\mathrm{SNR}=var(\mathbf{X}\beta)/\sigma_{e}^{2}\in\{0.5,\ 1,\ 2\}$. For all the above settings, we had $n=1,000$, $p=5,000$, $K=250$ with 20 variables in each group.
\begin{figure}[h!]
  \centering
    \includegraphics[height=6cm]{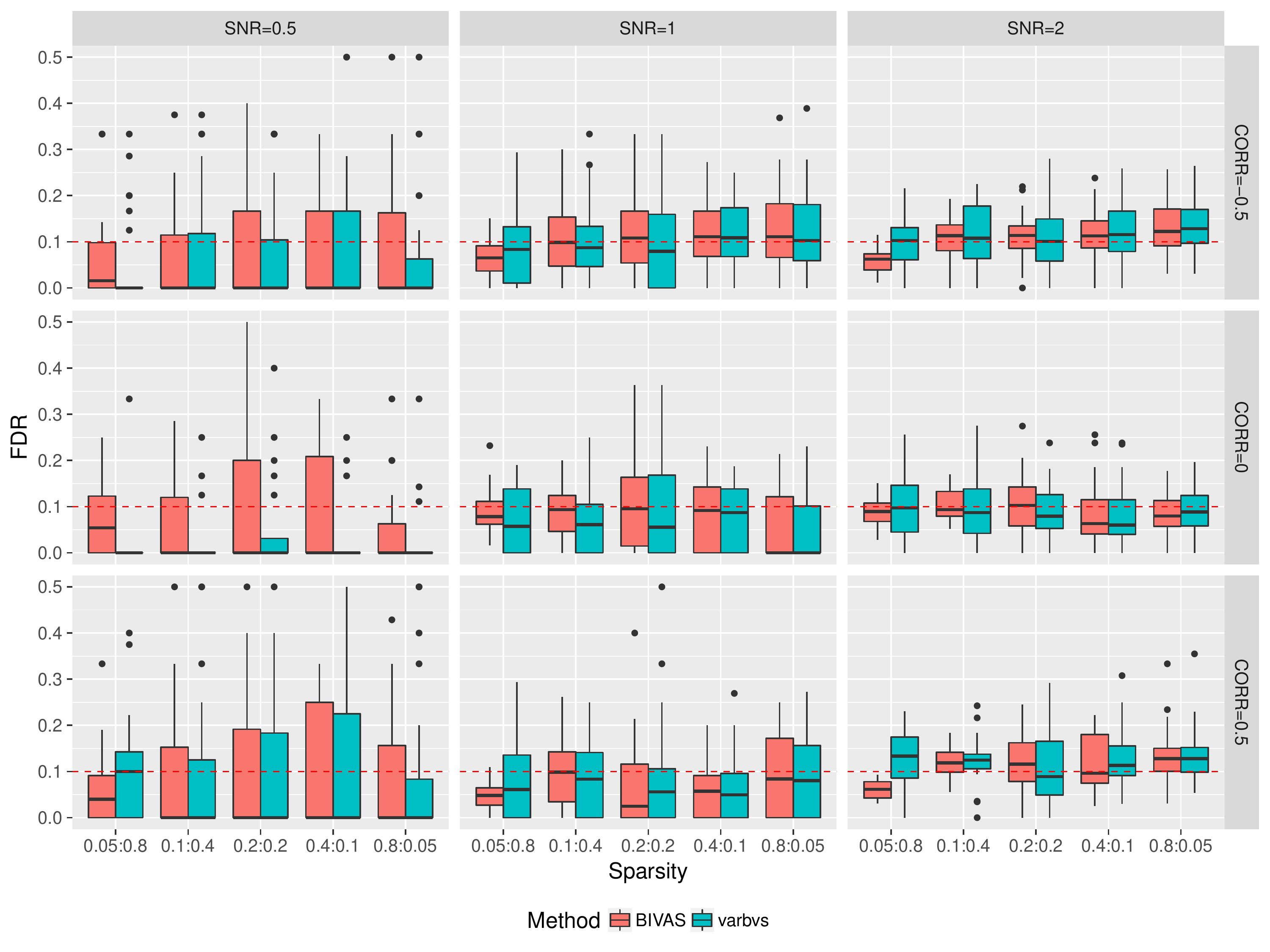}
    \includegraphics[height=6cm]{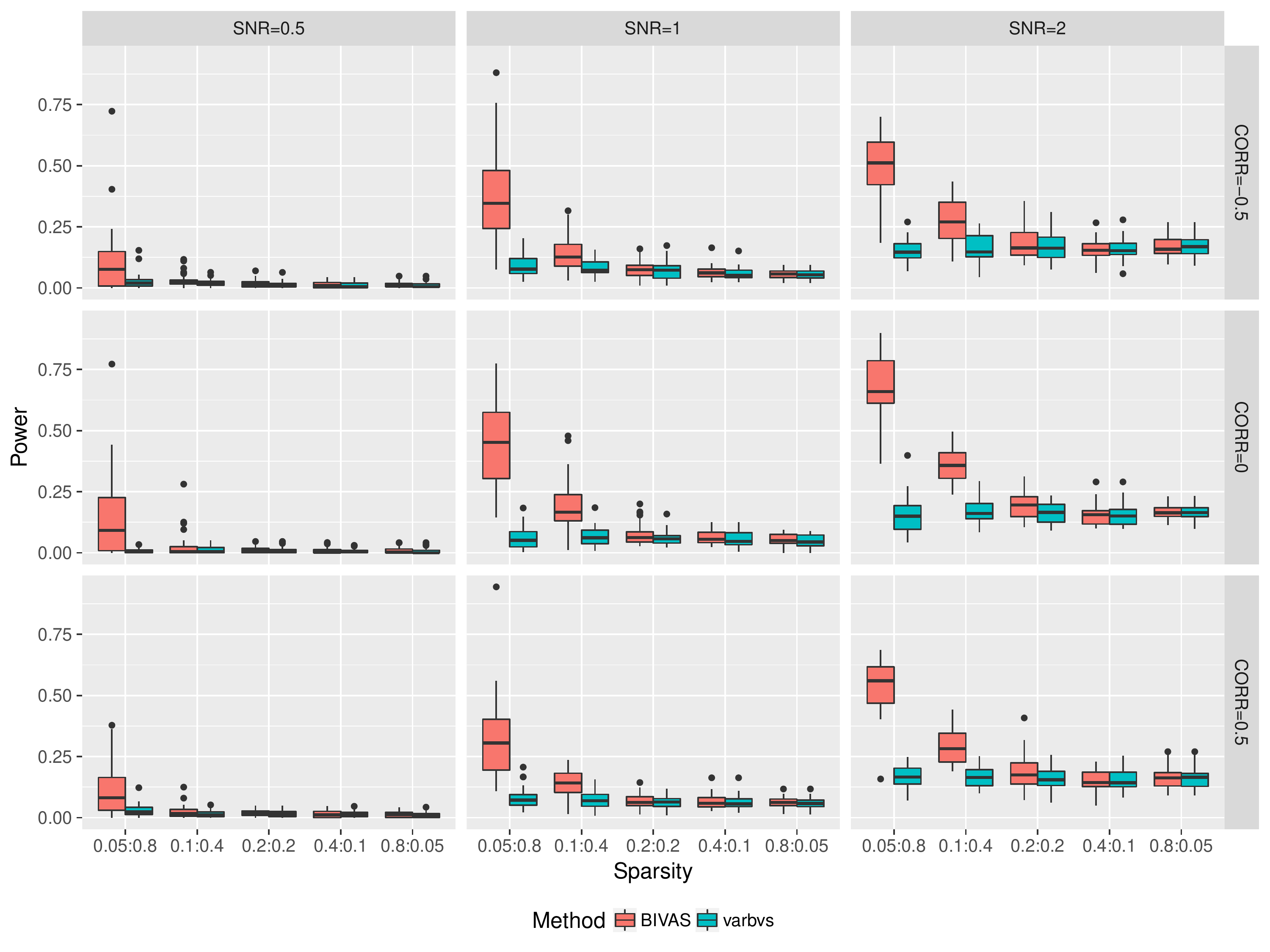}    
    \caption{Comparison of BIVAS and varbvs for individual variable selection.\label{fig1}}
    \centering
\end{figure}

As cMCP and GEL did not provide FDR estimates for variable selection, we first compared BIVAS with varbvs. Figure~\ref{fig1} shows the performance of FDR control and statistical power for individual variable selection obtained by BIVAS and varbvs. When the SNR is small, both methods are underpowered. However, BIVAS gains more power as the group sparsity dominates and further enlarges the gap as SNR increases. As $\rho$ moves away from zero, empirical FDRs of both methods are slightly inflated.
\begin{figure}[h!]
\centering
    \includegraphics[height=6cm]{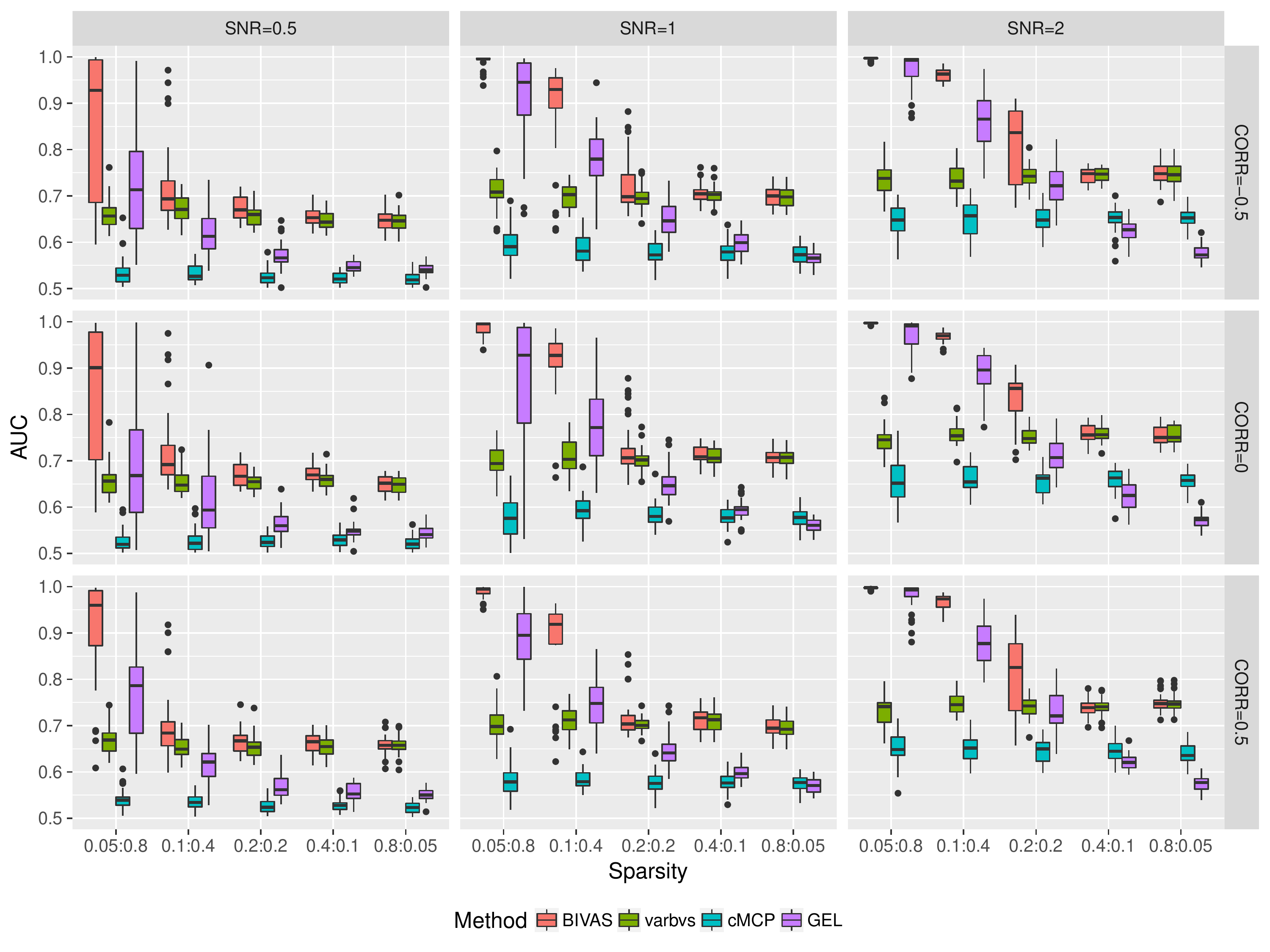}
    \includegraphics[height=6cm]{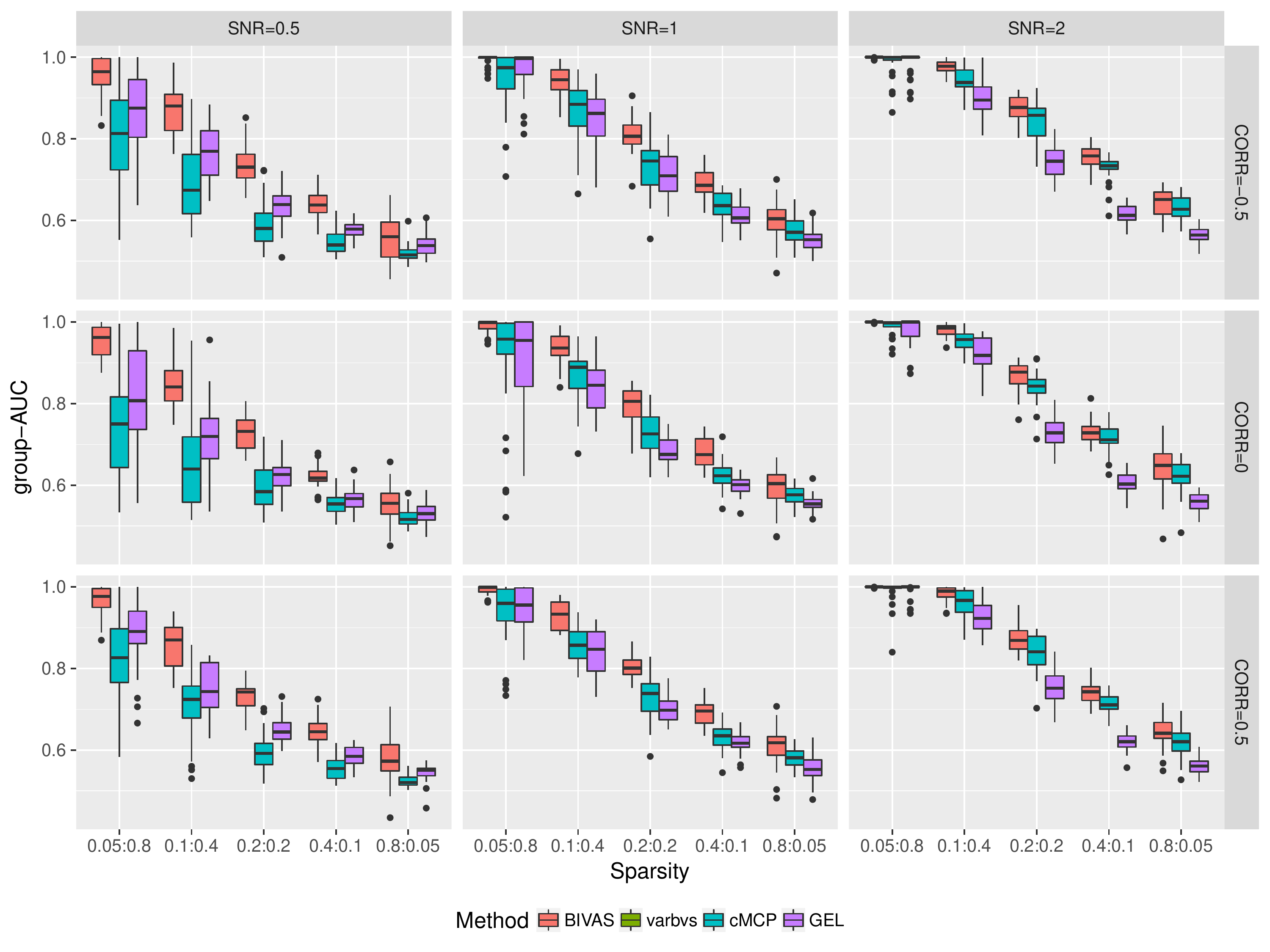}
    \includegraphics[height=6cm]{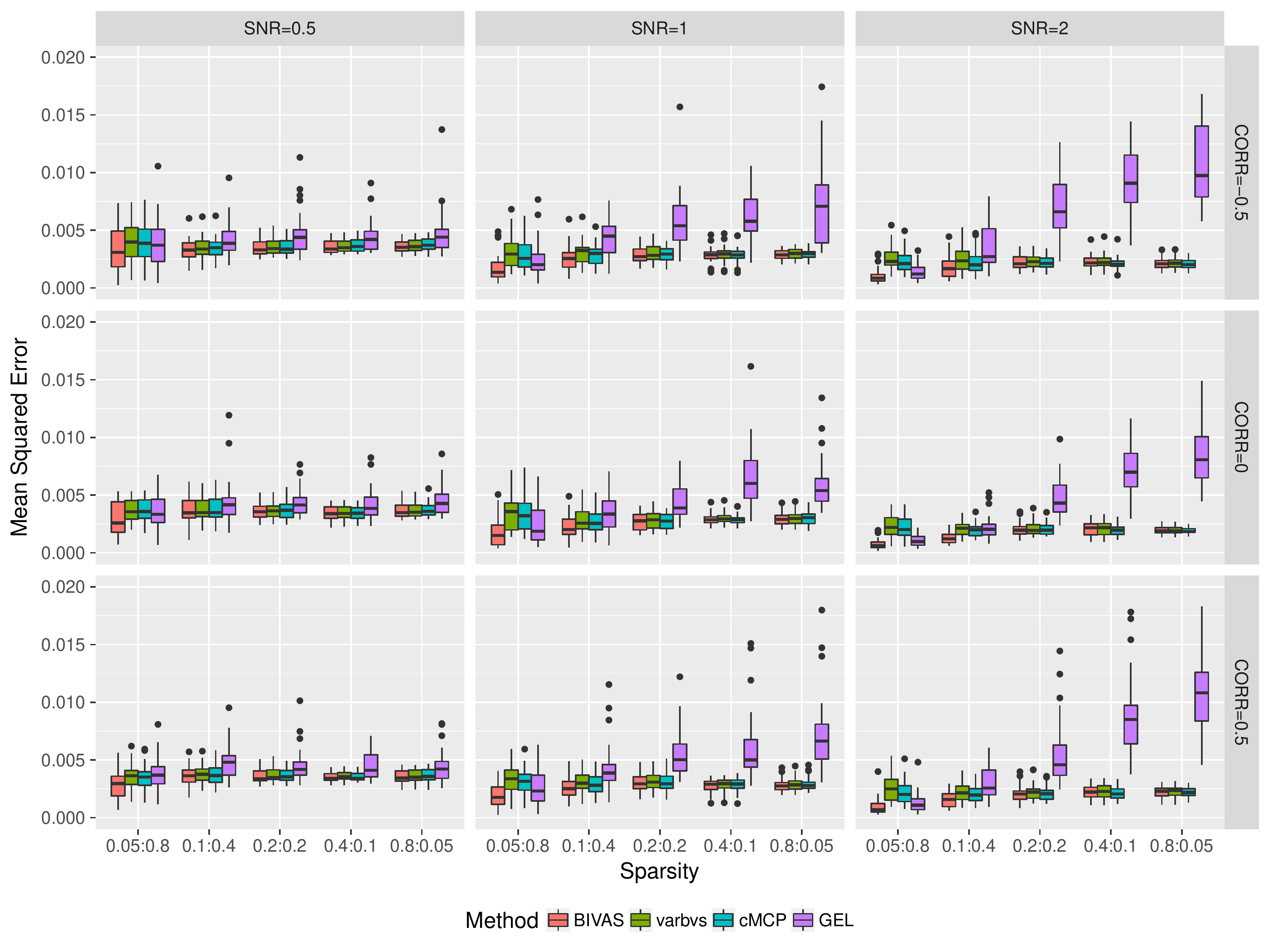}
    \includegraphics[height=6cm]{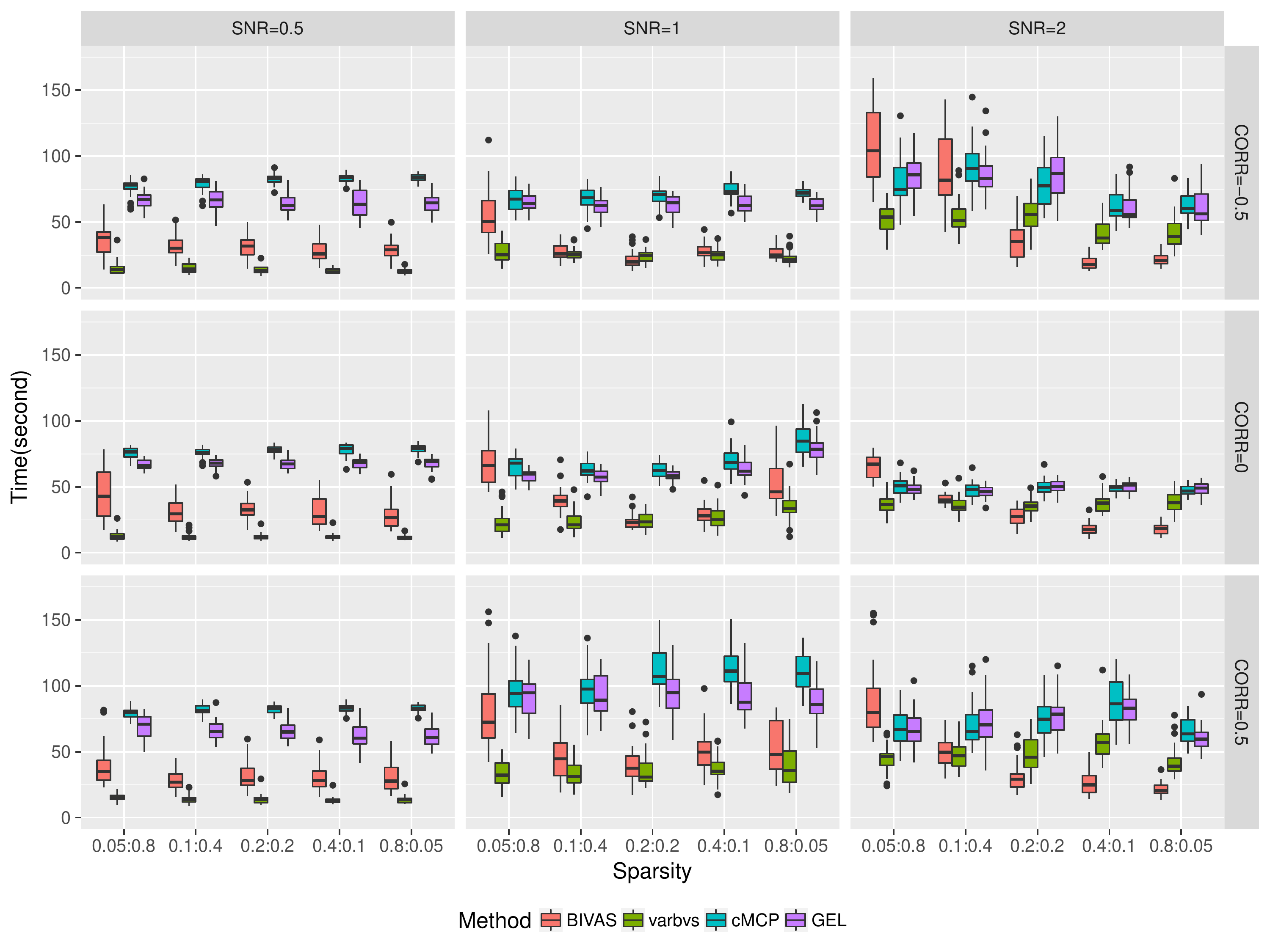}
    \caption{Comparisons of BIVAS, varbvs, cMCP and GEL (coupling parameter $\tau=1/3$). Top Left: AUC for individual variable selection. Top Right: AUC for group selection. Bottom Left: Mean squared error (MSE) of coefficient estimates. Bottom Right: Computational time.\label{fig2}}
    \centering
\end{figure}

Figure~\ref{fig2} shows the comparison of BIVAS with varbvs, cMCP and GEL in terms of bi-level variable selection, estimation accuracy and computational efficiency. As varbvs only selects individual variables, we treat it as a base line for comparisons of BIVAS with other two alternatives. In the bottom left penal, estimation errors of all the three methods decrease steadily as SNR increases when the sparsity-in-group dominates. BIVAS has similar performance with cMCP and GEL when SNR is moderate ($\mathrm{SNR}=0.5$) but outperforms them when SNR is relatively large ($\mathrm{SNR}=1, 2$). When sparsity-in-variable dominates, the estimation performances of BIVAS and cMCP are close to varbvs, but the estimation error of GEL is inflated.

To evaluate the performance of variable selection, we primarily focus on the measure of area under the receiver operating characteristic (ROC) curve (AUC) both at the group and individual levels. The top left panel of Figure~\ref{fig2} shows that the performance of variable selection for BIVAS is comparable with GEL when SNR is large. When the signal is weak ($\mathrm{SNR}=0.5$), the AUC of BIVAS is much larger than that of GEL. Moreover, as the `bulk' of sparsity moves to individual variable level, BIVAS converges to varbvs while GEL becomes even worse than varbvs. This pattern is consistent with that we observe in the measurement of estimation error. In all settings we considered, the perfomance of cMCP is poor.
The top right panel in Figure~\ref{fig2} shows the performance of variable selection at the group level (group-AUC). The pattern of group-AUC is similar to the individual level AUC. The bottom right panel in Figure~\ref{fig2} illustrates the computational efficiency of the four methods. With multi-thread computation, the speed of BIVAS is comparable to other methods and faster than cMCP and GEL in most cases.

\begin{figure}[h!] 
  \centering
    \includegraphics[height=7cm]{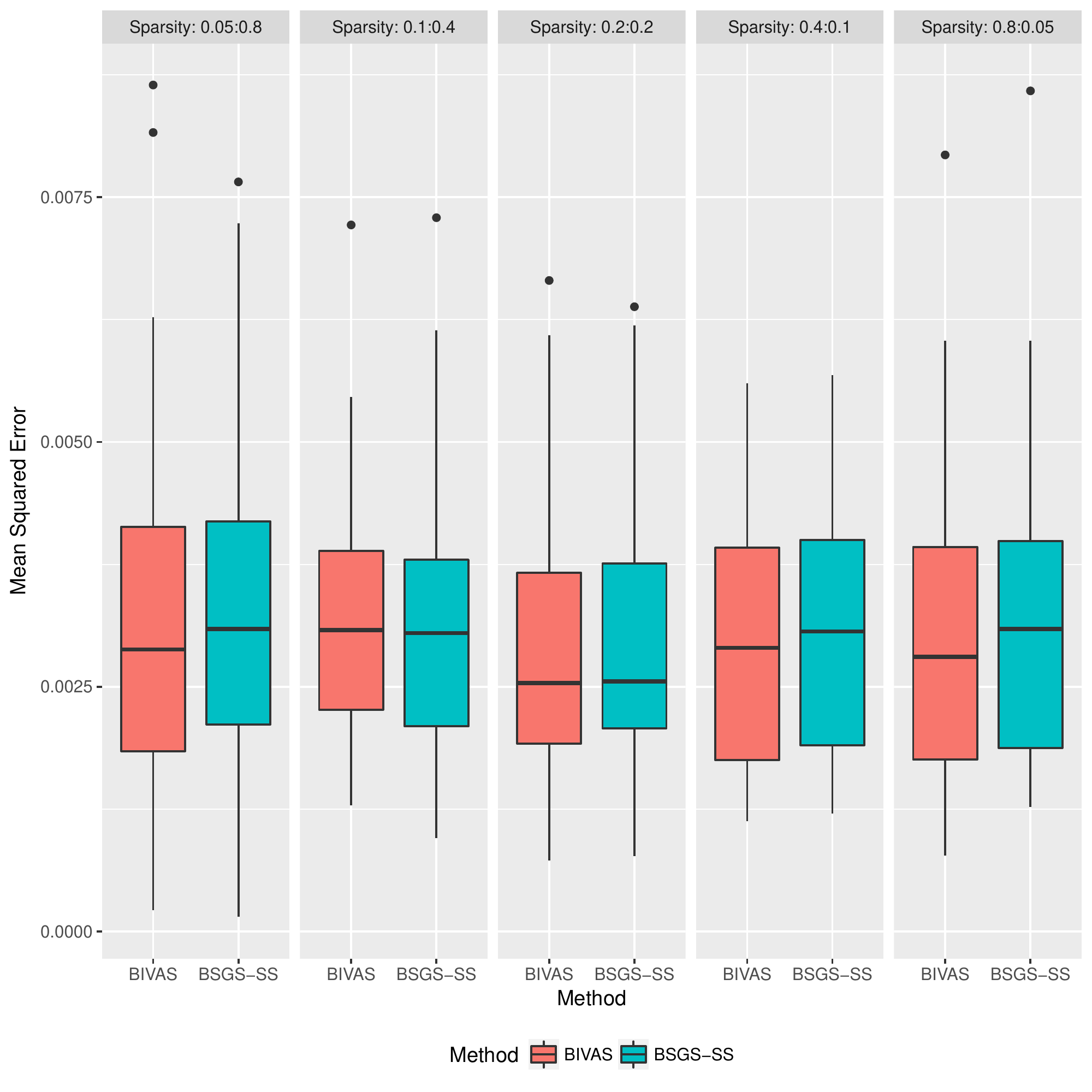}
    \includegraphics[height=7cm]{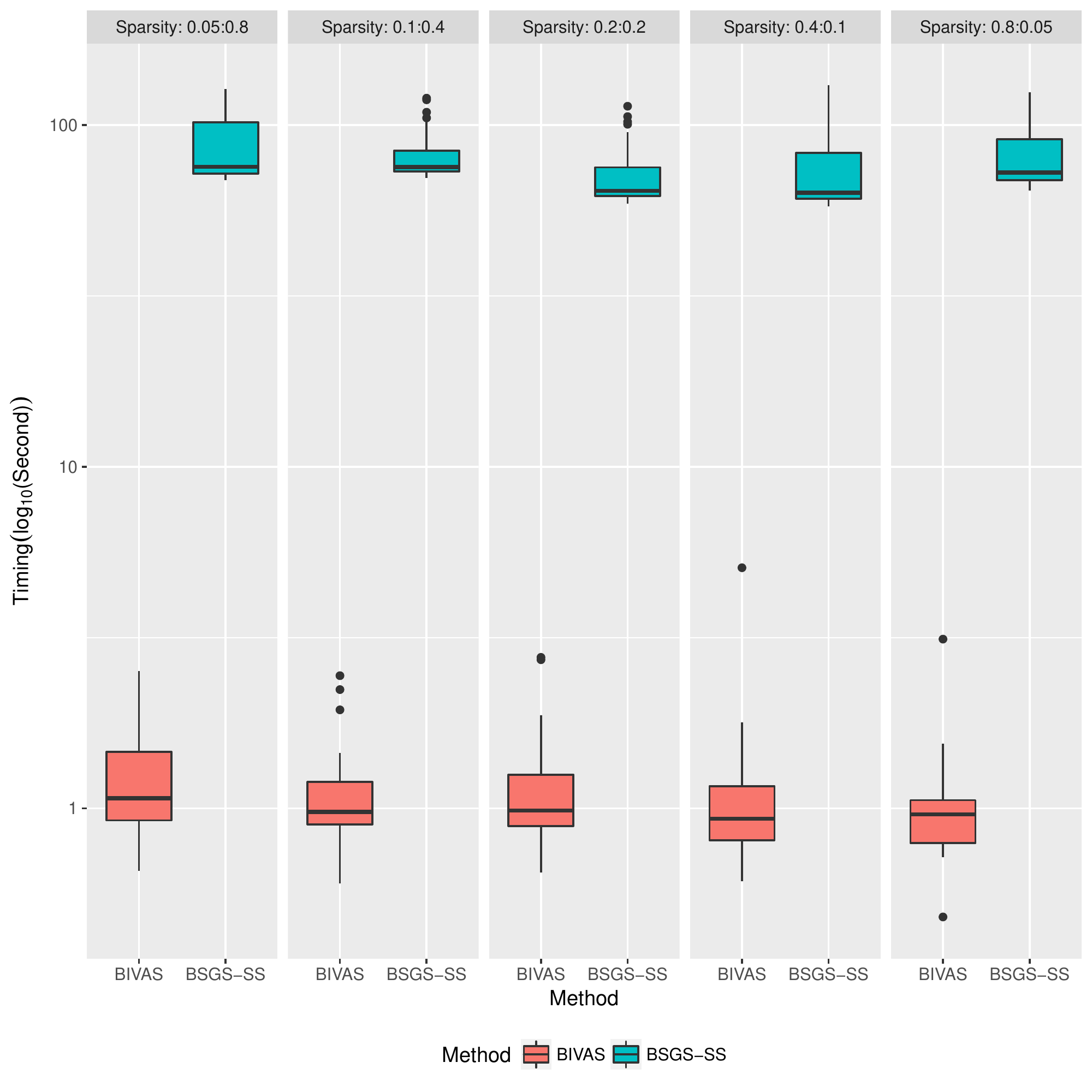}   
    \caption{Comparison of BIVAS and BSGS-SS. Left: Mean Squared Error of coefficient estimates. Right: Time.\label{fig3}}
    \centering
\end{figure}

In addition, we also made comparisons of the estimation accuracy and computational efficiency between BIVAS and Bayesian methods adopting MCMC. Here we considered BSGS-SS \citep{xu2015bayesian}; we set $n=200$, $p=1,000$, $K=100$ with $10$ variables in each group and $\rho=0.5$, $\mathrm{SNR}=1$. As illustrated in Figure~\ref{fig3}, BIVAS achieves almost the same estimation accuracy as BSGS-SS but uses only around $1\%$ of its computational time.

\begin{figure}[h!]
  \centering
    \includegraphics[height=8cm]{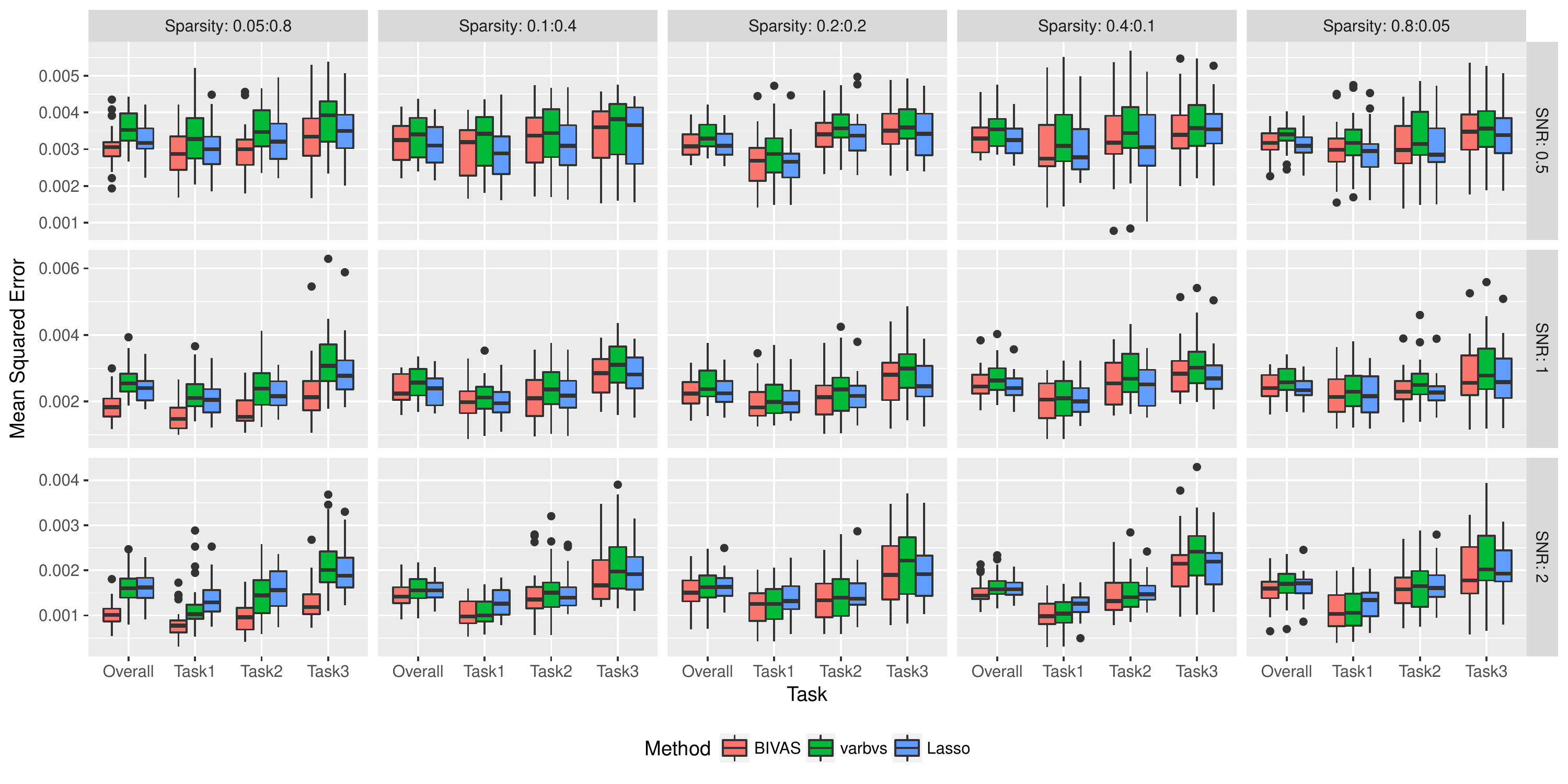}
    \caption{Comparison of BIVAS, varbvs, Ridge and Lasso in multi-task learning.\label{fig4}}
    \centering
\end{figure}

For multi-task BIVAS, we compared with varbvs and Lasso that are applied separately to each task. We simulated $L=3$ tasks with sample sizes $n_{1}=600,\ n_{2}=500,\ n_{3}=400$. Number of variables $K=2,000$ was used throughout. We followed the settings in group BIVAS for the sparsity pattern and SNR. The estimation error was evaluated on both overall scale and individual-task scale, as shown in Figure~\ref{fig4}.
As one can observe, BIVAS outperforms varbvs and Lasso when the group sparsity is predominant and the difference increases as the signal becomes stronger. Even when the proportion of group sparsity decreases, BIVAS is still comparable with the other two alternatives. In addition, when a strong group-sparsity pattern exists (leftmost column), BIVAS has its biggest gain on Task 3, which has the smallest sample size. This is because BIVAS takes the advantage of shared sparsity pattern in different tasks.
\subsection{Real Data Analysis}
To examine the performance of BIVAS in large scale data, we provide three real examples: we first apply the regression model to the GWAS data from the Wellcome Trust Case Control Consortium (WTCCC) \citep{wellcome2007genome} and the Northern Finland Birth Cohort (NFBC) \citep{sabatti2009genome}; then we analyze a movie review data set from IMDb.com \citep{maas2011learning} using the multi-task model.

\subsubsection{GWAS data}
In the GWAS data sets, we conducted quality control based on PLINK \citep{purcell2007plink} and GCTA \citep{yang2011gcta}: individuals with $>2\%$ missing genotypes were first removed; we also removed the SNPs with minor allele frequency $<0.05$, missingness $>1\%$, or p-value $<0.001$ in Hardy-Weinberg equilibrium test, excluding individuals with genetic relatedness greater than $0.025$.

\begin{figure}[h!] 
  \centering
    \begin{subfigure}[t]{0.24\textwidth}
        \centering
        \includegraphics[width=3.6cm]{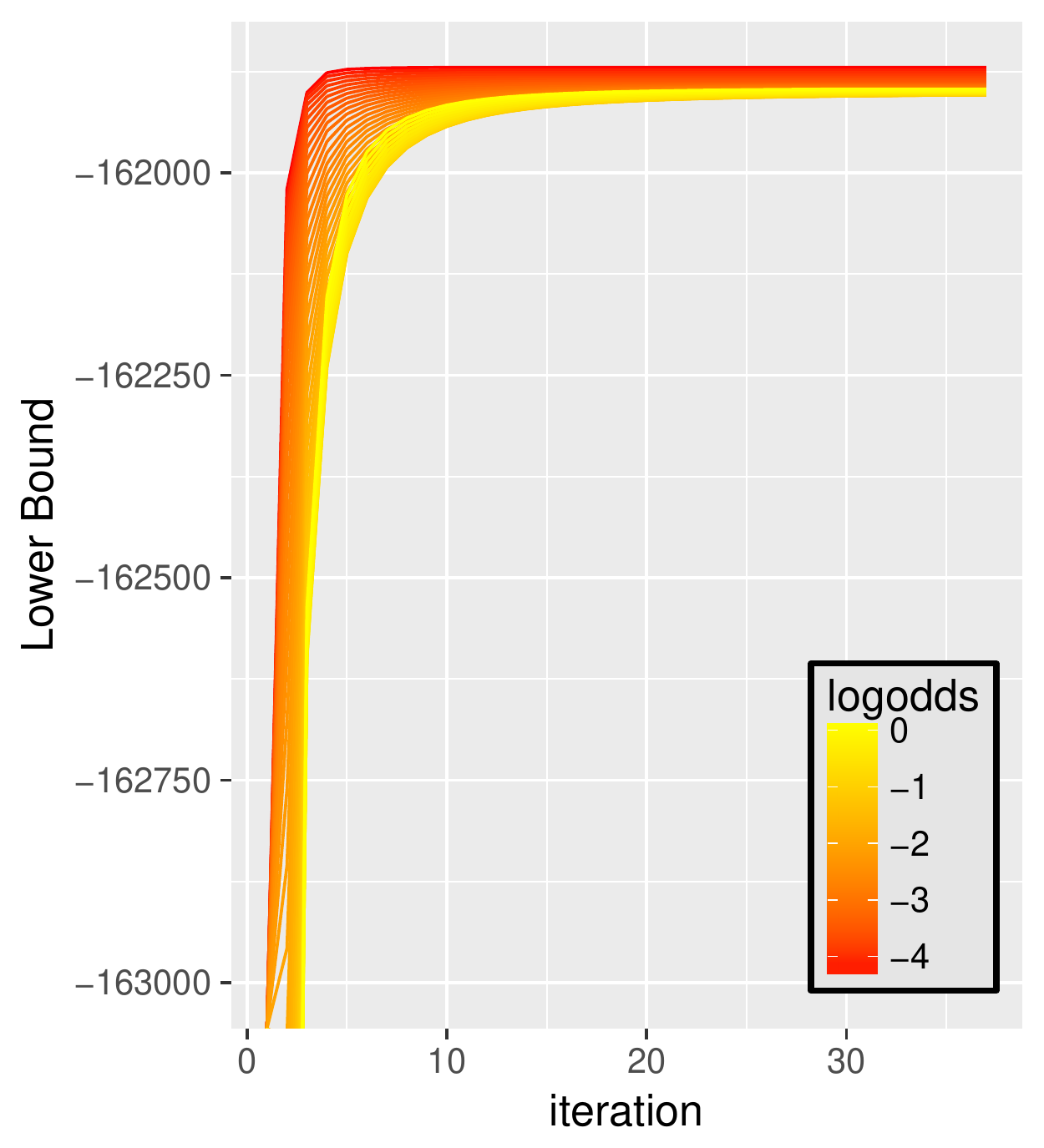}
        \caption{}
      \end{subfigure}%
    ~
    \begin{subfigure}[t]{0.24\textwidth}
        \centering
        \includegraphics[width=3.6cm]{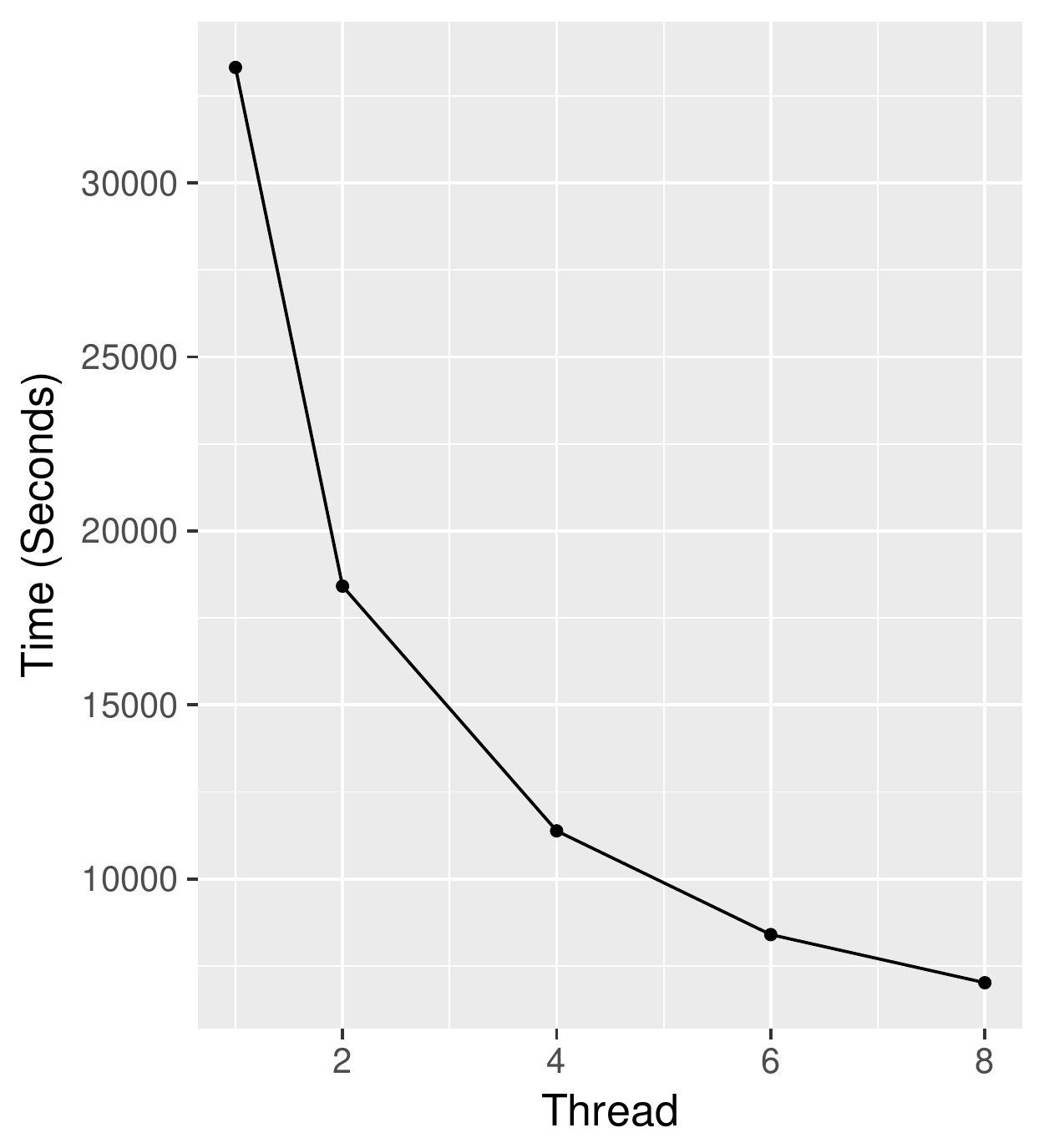}
        \caption{}
      \end{subfigure}%
    ~
    \begin{subfigure}[t]{0.24\textwidth}
        \centering
        \includegraphics[width=3.6cm]{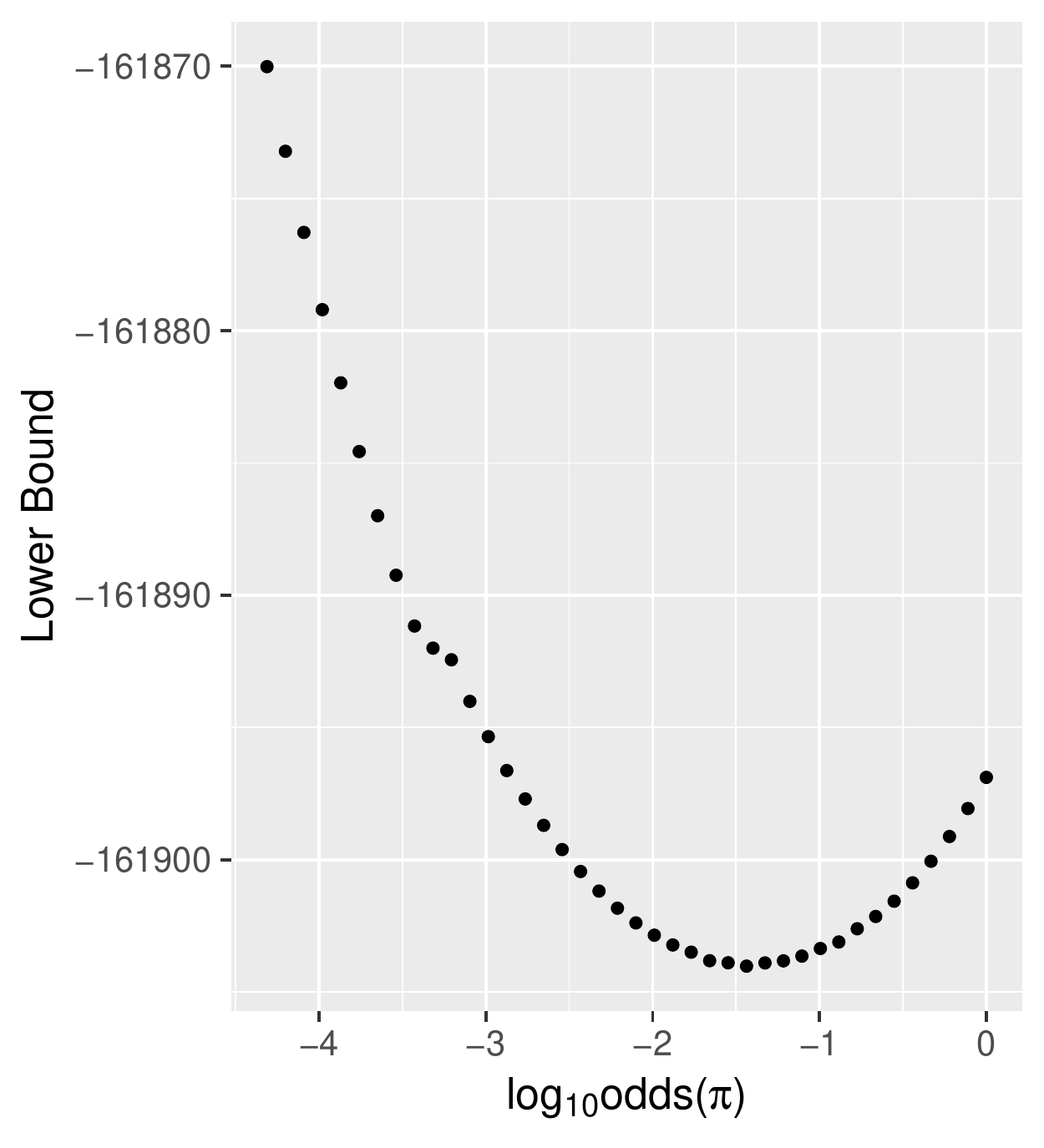} 
        \caption{}
      \end{subfigure}%
    ~
    \begin{subfigure}[t]{0.2\textwidth}
        \centering
        \includegraphics[width=3.6cm]{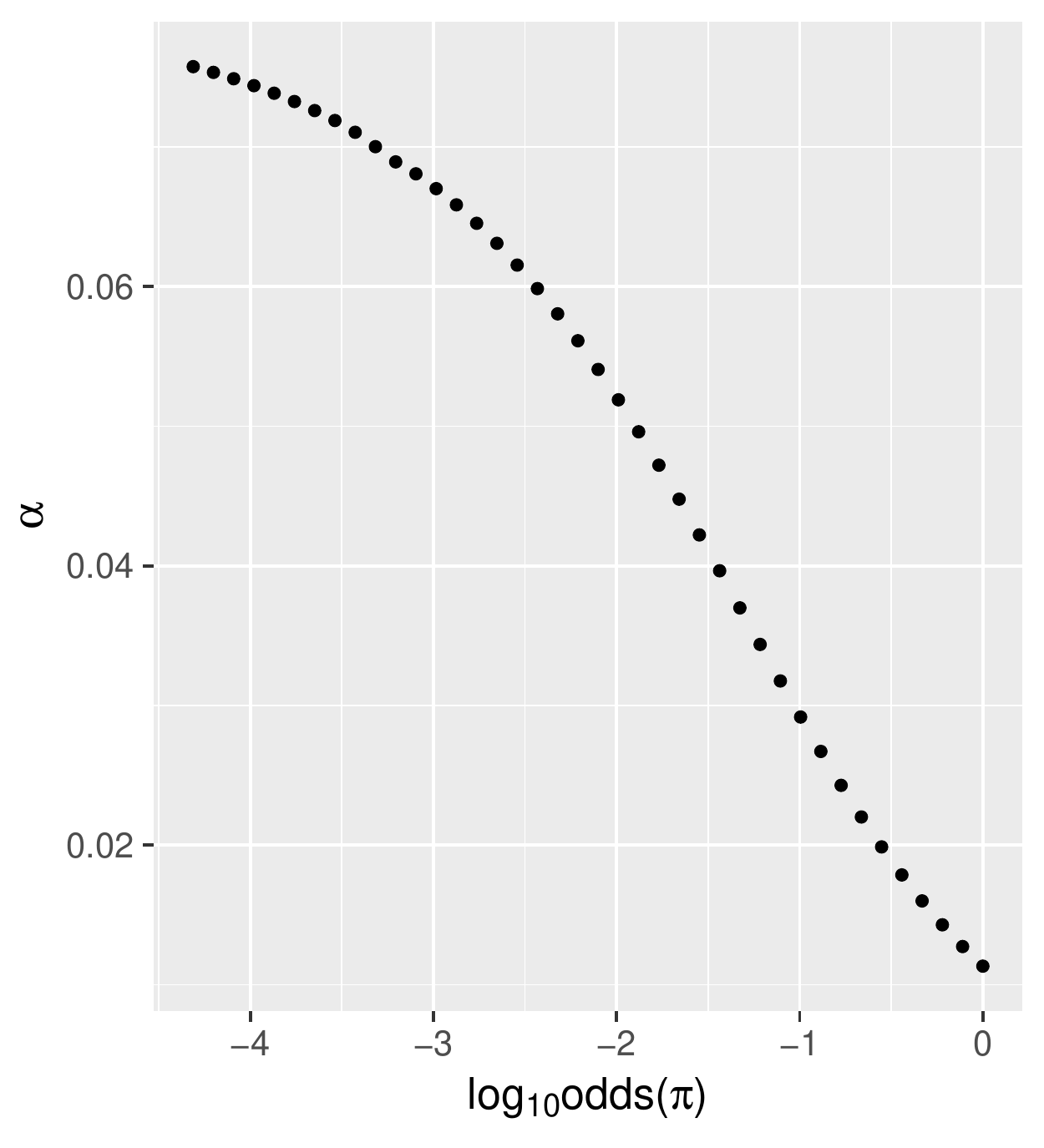}  
        \caption{}
      \end{subfigure}%
    \caption{BIVAS in fitting HDL. (a) Convergence of lower bound for $h=40$ EM procedure. (b) Computational times using 1, 2, 4, 6, 8 threads. (c) Lower bound for the $40$ settings procedure after convergence. (d) $\hat{\alpha}$ for the $40$ settings after convergence.\label{fig5}}
    \centering
\end{figure}

\begin{figure}[h!] 
  \centering
    \includegraphics[width=16cm]{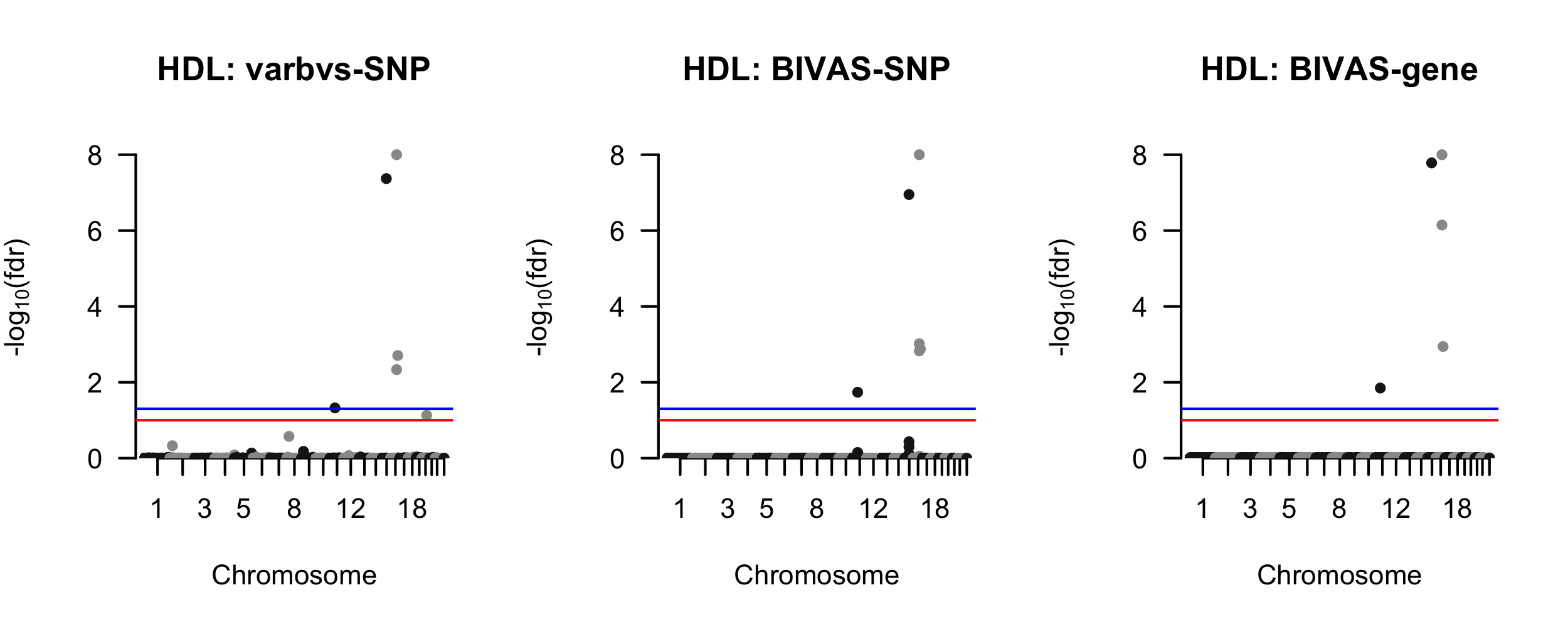}    
    \caption{Manhattan plots of High-Density Lipoprotein (HDL). Red line represents $fdr=0.1$ and blue line represents $fdr=0.05$.\label{fig6}}
    \centering
\end{figure}

We first considered the High-Density Lipoprotein (HDL) from the NFBC data, which was accessed by the database of Genotypes and Phenotypes (dbGaP) at \url{https://www.ncbi.nlm.nih.gov/projects/gap/cgi-bin/study.cgi?study_id=phs000276.v2.p1}. This data set was composed of $5,123$ individuals and $319,147$ SNPs. 
In our analysis, the SNPs were first annotated with their corresponding gene region using ANNOVAR \citep{doi:10.1093/nar/gkq603}, which leads to $318,686$ SNPs in $20,493$ genes without overlap. Treating the genes as groups, we applied both BIVAS and varbvs to the data. Figure~\ref{fig5} (a) shows the convergence of each EM procedure for BIVAS. One can observe that the EM algorithm converges faster for smaller values of $\pi$, suggesting the evidence of group sparsity. Computational times for different numbers of threads are presented in Figure~\ref{fig5} (b). When $h=40$, BIVAS took around 3.2 hours to converge using 4 threads and only took 1.6 hours using 8 threads, which indicates that the developed algorithm achieved almost perfect efficiency in parallelization. Estimates of lower bound and parameter $\alpha$ are shown in Figure~\ref{fig5} (c) and (d), suggesting the effectiveness of leveraging group structure using BIVAS. After the convergence, we identified the SNPs and genes based on $fdr<0.05$. Five risk variants (rs2167079, rs1532085, rs3764261, rs7499892, rs255052) were identified by varbvs. BIVAS discovered one more variant: rs1532624. For the group level selection, BIVAS identified five associated genes, among which \textit{CETP} containsed two risk SNPs: rs7499892 was also identified by varbvs but rs1532624 was a new one. The above results are visualized in the Manhattan plots (Figure~\ref{fig6}).

\begin{figure}[h!]
  \centering
    \includegraphics[width=16cm]{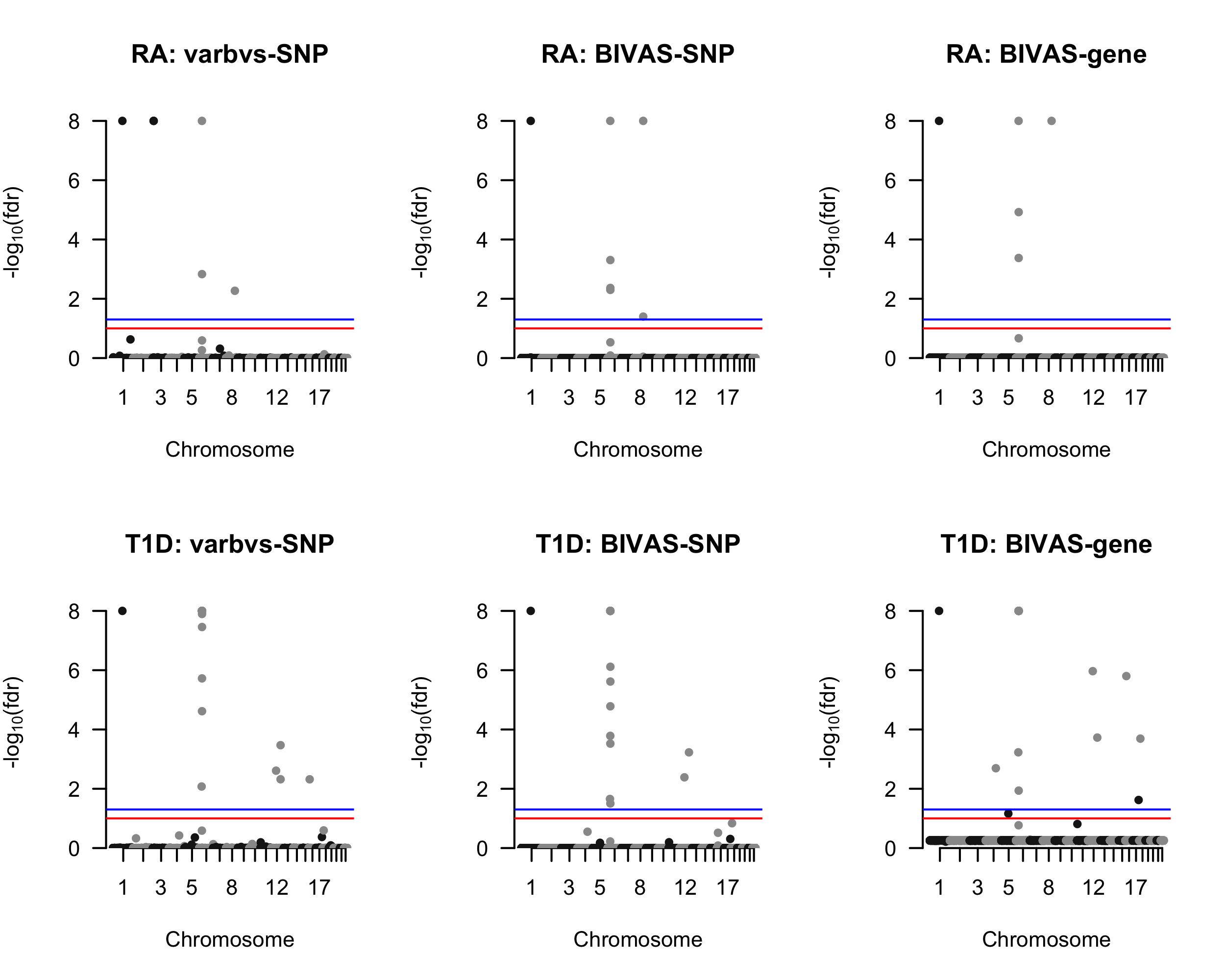}    
    \caption{Manhattan plots of Rheumatoid Arthritis (RA) and Type 1 Diabetes (T1D). Red line represents $fdr=0.1$ and blue line represents $fdr=0.05$.\label{fig7}}
    \centering
\end{figure}

In the second example, we analyzed Rheumatoid Arthritis (RA) and Type 1 Diabetes (T1D) in the WTCCC data. These data sets were from European Genome-phenome Archive (EGA) websites \url{http://www.ebi.ac.uk/ega/studies/EGAS00000000011} and \url{http://www.ebi.ac.uk/ega/studies/EGAS00000000014}. After quality control, we had $4,494$ individuals and $307,089$ SNPs for RA, and $4,986$ individuals and $307,357$ SNPs for T1D. The SNPs were then matched with corresponding genes using HapMap3 as reference, leading to $242,597$ SNPs with $16,789$ genes for RA and $242,824$ SNPs with $16,815$ genes for T1D. Manhattan plots are shown in Figure~\ref{fig7}. At the SNP level, the identification results of BIVAS and varbvs are similar but BIVAS further interrogated signals at the gene level making the results more interpretable. For example, in T1D, genes \textit{ADA1}, \textit{LINC00469} and \textit{LOC100996324} were identified as associated by BIVAS, but these genes contain no single associated SNP either identified by varbvs or BIVAS. This suggests that the associations are weak at the SNP level, but they aggregatively improve power as a group and hence identified by BIVAS at the gene level.

\subsubsection{IMDB movie data}
In the third example, we analyzed IMDb dataset \citep{maas2011learning} based on multi-task BIVAS. The IMDb data set was publicly availible at \url{IMDb.com}. The original data were extracted from movie reviews from IMDb.com. It contained $50K$ movie reviews that were equally split into a training set and a test set. Each review was marked with a rating ranging from 0 to 10, only the polarized reviews were retained (rating$>7$ or rating $<4$). The dataset was comprised of equal number of positive reviews and negative reviews. A bag of representative words was concluded from the whole review. Based on the bag of words, we adopted binary representation to indicate presence of the words. This led to $K=27,743$ features (words) with the rating being the response variable. We used 6 genres of movies as our tasks: drama, comedy, horror, action, thriller and romance. Only the reviews of movies that had exactly one genre were used. This led to the sample sizes $3,354$ for drama, $2,235$ for comedy, $1,175$ for horror, 346 for action, 258 for thriller and 139 for romance. We compared BIVAS against Ridge, Lasso and varbvs.

Table~\ref{tab1} shows the testing errors of the four methods. For the categories of Horror, Action, Thriller and Romance, BIVAS has better performance than the other 3 methods. Note that these genres have smaller sample sizes compared to comedy and action. This result is consistent with what we obtain in the simulation study.
\begin{table}[ht]
\centering
\begin{tabular}{rrrrrrrr}
  \hline
 & Overall & Drama & Comedy & Horror & Action & Thriller & Romance \\ 
  \hline
ridge & 9.58 & 9.01 & 10.55 & 9.01 & 10.99 & 10.14 & 6.76 \\ 
  lasso & 6.67 & \textbf{6.13} & \textbf{6.67} & 7.20 & 8.65 & 9.27 & 6.77 \\ 
  varbvs & 7.14 & 6.20 & 6.90 & 8.94 & 8.37 & 11.48 & 6.91 \\ 
  BIVAS & \textbf{6.66} & 6.32 & 7.01 & \textbf{6.76} & \textbf{6.89} & \textbf{7.44} & \textbf{5.39} \\ 
   \hline
\end{tabular}
\caption{IMDb testing error.\label{tab1}} 
\label{t:}
\end{table}

\begin{figure}[h!]
  \centering
    \includegraphics[height=11.5cm]{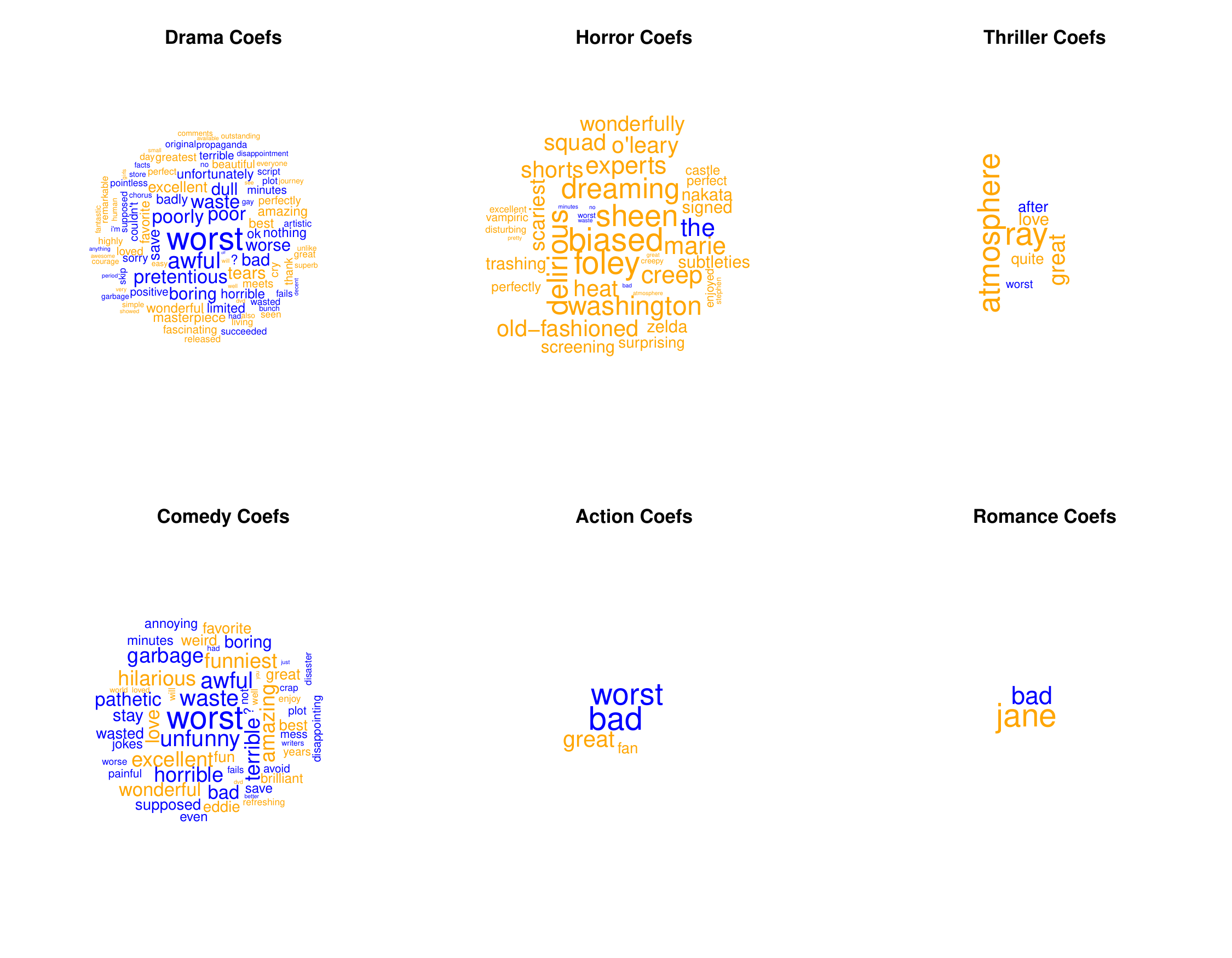}
    \caption{IMDb wordcloud generated by varbvs.\label{fig8}}
    \centering
\end{figure}

\begin{figure}[h!]
  \centering
    \includegraphics[height=11.5cm]{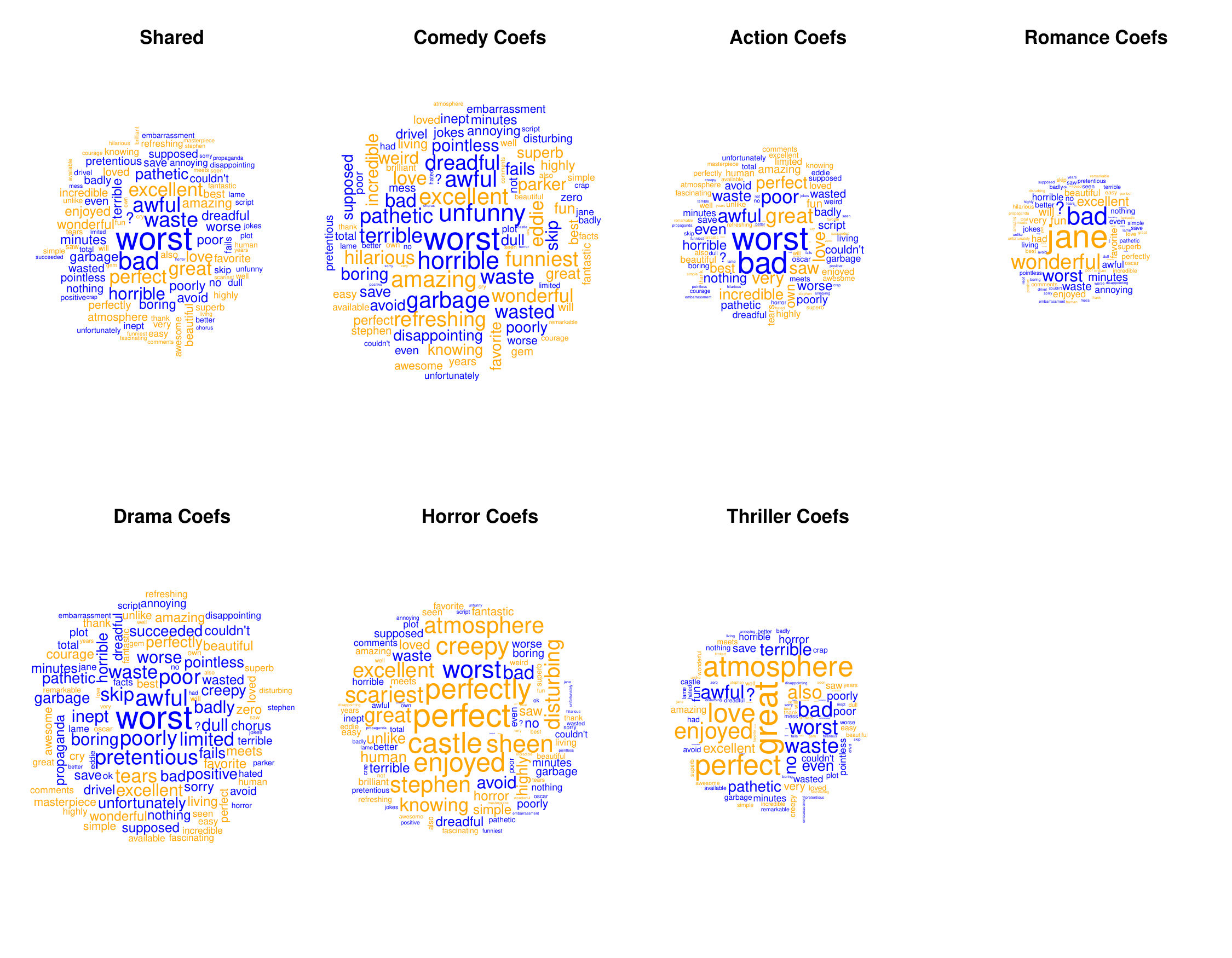}
    \caption{IMDb wordcloud generated by BIVAS.\label{fig9}}
    \centering
\end{figure}

The words selected by BIVAS and varbvs are presented in Figure~\ref{fig8} and Figure~\ref{fig9} using `wordcloud' package in R. The words in blue and yellow represent the negative and positive effects, respectively. The size of words represents the effect size. As shown in Figure~\ref{fig8}, small number of words were identified by varbvs to be associated with Action, Horror or Thriller, which are genres with smallest sample sizes. However, as shown in Figure~\ref{fig9}, BIVAS greatly enriches the effective words in these tasks by borrowing information from the large samples (Drama, Comedy and Horror). Many associated words that were overwhelmed by noise are now revealed. This can be viewed as a consequence of bi-level selection which selects the important variables and, at the same time, allows sparsity pattern to be shared within group (or through tasks in multi-task learning). Hence, many useful words shared through tasks, such as `worst', `awful' and `amazing', can be revealed for small sample and some particular predictors, like `scariest' in Thriller and Horror, are maintained task-specific. On the other hand, varbvs (as well as Ridge and Lasso) does not account for the bi-level sparsity structure, so it is unable to capture the shared information through tasks.

\section{Discussion}
The bi-level variable selection aims at capturing the sparsity at both the individual variable level and the group level to better interrogate the structural information that can assist parameter estimation and variable selection. Bayesian bi-level selection methods are free of parameter tuning and able to obtain the posterior distributions of random effects. Based on the posterior distributions, variables can be selected at both levels by controlling $fdr$. Despite the convenience, existing Bayesian bi-level variable selection methods are often computationally inefficient and unscalable to large data sets due to the intractable posterior.

In this paper, we propose a hierarchically factorizable formulation to approximate the posterior distribution, by utilizing the structure of bi-level variable selection. Under the variational assumption, a computationally efficient algorithm is developed based on the variational EM algorithm and importance sampling. The convergence of algorithm is promised and the accurate approximation for the posterior mean can be obtained. The proposed algorithm is efficient, stable and scalable. Our software is fast and capable of parallel computing. After convergence, variable selection at both levels can be conducted by controlling the $fdr$, prediction can be made based on posterior means. Through the simulation study we showed that our method is no worse than alternative methods given the same computational cost and outperforms some methods in many cases. We also applied BIVAS to real world data and verified its scalability and capability of bi-level selction.

\bigskip
\begin{center}
{\large\bf SUPPLEMENTARY MATERIAL}
\end{center}

\begin{description}

\item[Supp\_final.pdf] The supplemental materials include the detailed derivation for both regression and multi-task learning. (PDF)

\item[R-package for BIVAS:] R-package `bivas'. The package contains the functions used in fitting BIVAS and making statistical inference. (zipped tar file)

\item[sim-bivas] The file contains the R scripts for generating numerical results in Section 3. (zipped file for R scripts)

\end{description}

\bibliographystyle{chicago}
\bibliography{ref}

\newpage
\renewcommand{\thesubsection}{\Alph{subsection}}
\section*{Supplementary Document}

\subsection{Variational EM Algorithm: Regression with BIVAS}

\subsubsection{E-Step}

Let $\bm{\theta}=\{\alpha,\pi,\sigma_{\beta}^{2},\sigma_{e}^{2},\bm{\omega}\}$ be the collection of model parameters in the main text. The joint probabilistic model is
\begin{eqnarray}
  \begin{aligned}
  &\mathrm{Pr}(\mathbf{y},\bm{\eta},\bm{\gamma},\bm{\beta}|\mathbf{X},\mathbf{Z};\bm{\theta}) 
  =\mathrm{Pr}(\mathbf{y}|\bm{\eta},\bm{\gamma},\bm{\beta},\mathbf{X},\mathbf{Z},\bm{\theta})\mathrm{Pr}(\bm{\eta},\bm{\gamma},\bm{\beta}|\bm{\theta})  \\
   =&\mathcal{N}(\mathbf{y}|\mathbf{Z}\bm{\omega}+\sum_{k}^{K}\sum_{j}^{l_{k}}\eta_{k}\gamma_{jk}\beta{jk}\mathbf{x}_{jk})\prod_{k=1}^{K}\pi^{\eta_{k}}(1-\pi)^{1-\eta_{k}}\prod_{j=1}^{l_{k}}\mathcal{N}(0,\sigma_{\beta}^{2})\alpha^{\gamma_{jk}}(1-\alpha)^{1-\gamma_{jk}}.
  \end{aligned}
\end{eqnarray}
The logarithm of the marginal likelihood is
\begin{eqnarray}
  \begin{aligned}
  \log\ p(\mathbf{y}|\mathbf{X},\mathbf{Z};\bm{\theta})&=\log\sum_{\bm{\gamma}}\sum_{\bm{\eta}}\int_{\bm{\beta}}\mathrm{Pr}(\mathbf{y},\bm{\eta},\bm{\gamma},\bm{\beta}|\mathbf{X},\mathbf{Z};\bm{\theta})d\bm{\beta}\\
  &\geq\sum_{\bm{\gamma}}\sum_{\bm{\eta}}\int_{\bm{\beta}}q(\bm{\eta},\bm{\gamma},\bm{\beta})\log \frac{\mathrm{Pr}(\mathbf{y},\bm{\eta},\bm{\gamma},\bm{\beta}|\mathbf{X},\mathbf{Z};\bm{\theta})}{q(\bm{\eta},\bm{\gamma},\bm{\beta})}d\bm{\beta}\\
  &=\mathbb{E}_q[\log\mathrm{Pr}(\mathbf{y},\bm{\eta},\bm{\gamma},\bm{\beta}|\mathbf{X},\mathbf{Z};\bm{\theta})-\log q(\bm{\eta},\bm{\gamma},\bm{\beta})]\\
  &\equiv \mathcal{L}(q),
  \end{aligned}
\end{eqnarray}
where we have adopted Jensen's inequality to obtain the lower bound $\mathcal{L}(q)$. Next step is to iteratively maximize $\mathcal{L}(q)$ instead of working with the marginal likelihood directly. As in the main text, we use the following hierarchically factorized distribution to approximate the true posterior:
\begin{equation}
q(\bm{\eta},\bm{\gamma},\bm{\beta})=\prod_{k}^{K}\left(q(\eta_{k})\prod_{j}^{l_{k}}(q(\beta_{jk}|\eta_{k},\gamma_{jk})q(\gamma_{jk}))\right),
\end{equation}
where we have assumed that groups are independent; and given a group, the factors inside are also independent. With this assumption, we first rewrite the ELBO as:
\begin{eqnarray}
  \begin{aligned}
    \mathcal{L}(q)&=\mathbb{E}_{q(\eta)}\left[\mathbb{E}_{q(\gamma,\beta|\eta)}\left[ \log \Pr(\mathbf{y},\bm{\eta},\bm{\gamma},\bm{\beta})-\log q(\bm{\eta},\bm{\gamma},\bm{\beta}) \right]\right].
  \end{aligned}
\end{eqnarray}
Let $q(\bfgamma_k)= \prod_j^{l_k}q(\gamma_{jk})$, $q(\bfbeta_k|\eta_k,\bfgamma_k) = \prod_{j}^{l_{k}}(q(\beta_{jk}|\eta_{k},\gamma_{jk})q(\gamma_{jk}))$ and \\
$q(\eta_k,\bfgamma_k,\bfbeta_k)=q(\eta_k)\prod_j^{l_k}q(\beta_{jk}|\eta_k,\gamma_{jk})q(\gamma_{jk})$, the lower bound can be written in the following form:
\scriptsize
  \begin{align} 
    &\mathcal{L}(q)\nonumber\\
    =& \sum_{\bm{\eta}}\prod_{k}^K q(\eta_k)\sum_{\bm{\gamma}}\prod_{k}^K q(\bfgamma_{k})\int_{\bm{\beta}}\left(\log\Pr(\mathbf{y},\bm{\eta},\bm{\gamma},\bm{\beta}) - \sum_{k}^K\log q(\eta_k,\bfgamma_k,\bfbeta_k)\right) \prod_{k}^K q(\bfbeta_{k}|\eta_k,\bfgamma_{jk})d\bm{\beta}\nonumber\\
    =& \sum_{\eta_k}q(\eta_k)\sum_{\bfgamma_k}\prod_{j}^{l_k}q(\gamma_{jk})\int\prod_{j}^{l_k}q(\beta_{jk}|\eta_k,\gamma_{jk})\left[ \sum_{\eta_{-k}}\prod_{k'\neq k}q(\eta_{k'})\sum_{\bfgamma_{-k}}\prod_{k'\neq k}q(\bfgamma_{k'})\int\log \Pr(\mathbf{y},\bm{\eta},\bm{\gamma},\bm{\beta})\prod_{k'\neq k}q(\bfbeta_k|\eta_k,\bfgamma_k)d\bfbeta_{k'} \right]d\bfbeta_k\nonumber\\
    &- \sum_{\eta_k}q(\eta_k)\sum_{\bfgamma_k}\prod_{j}^{l_k}q(\gamma_{jk})\int\prod_{j}^{l_k}q(\beta_{jk}|\eta_k,\gamma_{jk})\log q(\eta_k,\bfgamma_k,\bfbeta_k)d\bfbeta_k + \mathrm{const}\nonumber\\
    =& \mathbb{E}_{q(\eta_k,\bfgamma_k,\bfbeta_k)}\left[ \mathbb{E}_{k'\neq k} \log\Pr(\mathbf{y},\bm{\eta},\bm{\gamma},\bm{\beta}) - \log q(\eta_k,\bfgamma_k,\bfbeta_k) \right] + \mathrm{const}\nonumber\\
    =&\mathbb{E}_{q(\eta_k)}\left[\mathbb{E}_{q(\bfgamma_k,\bfbeta_k|\eta_k)}\left[\mathbb{E}_{k'\neq k}\log\Pr(\mathbf{y},\bm{\eta},\bm{\gamma},\bm{\beta}) - \log q(\eta_k,\bfgamma_k,\bfbeta_k) \right]\right] + \mathrm{const}\nonumber\\
    =&q(\eta_k=1)\left[\mathbb{E}_{q(\bfgamma_k,\bfbeta_k|\eta_k=1)}\left[\mathbb{E}_{k'\neq k}\log\Pr(\mathbf{y},\bm{\eta}_{-k},\eta_k=1,\bm{\gamma},\bm{\beta}) - \log q(\eta_k=1,\bfgamma_k,\bfbeta_k) \right]\right]\\
    &+q(\eta_k=0)\left[\mathbb{E}_{q(\bfgamma_k,\bfbeta_k|\eta_k=0)}\left[\mathbb{E}_{k'\neq k}\log\Pr(\mathbf{y},\bm{\eta}_{-k},\eta_k=0,\bm{\gamma},\bm{\beta}) - \log q(\eta_k=0,\bfgamma_k,\bfbeta_k) \right]\right] + \mathrm{const},\nonumber
  \end{align}
\normalsize
where $\eta_k$ is from Bernoulli distribution and $\bfeta_{-k}$ is a vector obtained by removing the $k$-th term from $\bfeta$. $\mathbb{E}_{k'\neq k}(\cdot)$ denotes taking expectation with respect to the terms outside the $k$-th group. Now given $q(\eta_k)$, when $\eta_k=1$, we can focus on the expectations in Equation (5):
\scriptsize
\begin{align}
  &\mathbb{E}_{q(\bfgamma_k,\bfbeta_k|\eta_k=1)}\left[\mathbb{E}_{k'\neq k}\log\Pr(\mathbf{y},\bm{\eta}_{-k},\eta_k=1,\bm{\gamma},\bm{\beta}) - \log q(\eta_k=1,\bfgamma_k,\bfbeta_k) \right]\nonumber\\
  =& \sum_{\bfgamma_k}\prod_{j}^{l_k}q(\gamma_{jk})\int_{\bfbeta_k}\left(\mathbb{E}_{k'\neq k}\log\Pr(\mathbf{y},\bm{\eta}_{-k},\eta_k=1,\bm{\gamma},\bm{\beta}) - \log q(\eta_k=1,\bfgamma_k,\bfbeta_k)\right)\prod_{j}^{l_k}q(\beta_{jk}|\eta_k,\gamma_{jk})d\bfbeta_k\nonumber\\
  =& \sum_{\bfgamma_k}\prod_{j}^{l_k}q(\gamma_{jk})\int_{\bfbeta_k}\left(\mathbb{E}_{k'\neq k}\log\Pr(\mathbf{y},\bm{\eta}_{-k},\eta_k=1,\bm{\gamma},\bm{\beta}) - \log q(\bfgamma_k,\bfbeta_k|\eta_k=1)\right)\prod_{j}^{l_k}q(\beta_{jk}|\eta_k,\gamma_{jk})d\bfbeta_k + \mathrm{const}\nonumber\\
  =&\sum_{\gamma_{jk}}q(\gamma_{jk})\int q(\beta_{jk}|\gamma_{jk},\eta_k=1)\left[\sum_{\gamma_{-j|k}}\prod_{j'\neq j|k}q(\gamma_{j'k})\int\mathbb{E}_{k'\neq k}\left[\log\Pr(\mathbf{y},\bm{\eta}_{-k},\eta_k=1,\bm{\gamma},\bm{\beta})\right]\prod_{j'\neq j|k}q(\beta_{j'k},\gamma_{j'k}|\eta_k=1)d\beta_{j'k}\right]d\beta_{jk}\nonumber\\
  &-\sum_{\gamma_{jk}}q(\gamma_{jk})\int q(\beta_{jk}|\gamma_{jk},\eta_k=1)\log q(\beta_{jk},\gamma_{jk}|\eta_k=1)d\beta_{jk} + \mathrm{const}\nonumber\\
  =& \mathbb{E}_{q(\beta_{jk},\gamma_{jk}|\eta_k=1)}\left[\mathbb{E}_{j'\neq j|k}\left[ \mathbb{E}_{k'\neq k}\log\Pr(\mathbf{y},\bm{\eta}_{-k},\eta_k=1,\bm{\gamma},\bm{\beta})\right] - \log q(\beta_{jk},\gamma_{jk}|\eta_k=1) \right]\nonumber + \mathrm{const}\\
  =&q(\gamma_{jk}=1)\mathbb{E}_{q(\beta_{jk}|\eta_k=1,\gamma_{jk}=1)}\left[\mathbb{E}_{j'\neq j|k}\left[ \mathbb{E}_{k'\neq k}\log\Pr(\mathbf{y},\bm{\eta}_{-k},\eta_k=1,\bm{\gamma}_{-jk},\gamma_{jk}=1,\bm{\beta})\right] - \log q(\beta_{jk},\gamma_{jk}=1|\eta_k=1) \right]\\
  &+q(\gamma_{jk}=0)\mathbb{E}_{q(\beta_{jk}|\eta_k=1,\gamma_{jk}=0)}\left[\mathbb{E}_{j'\neq j|k}\left[ \mathbb{E}_{k'\neq k}\log\Pr(\mathbf{y},\bm{\eta}_{-k},\eta_k=1,\bm{\gamma}_{-jk},\gamma_{jk}=0,\bm{\beta})\right] - \log q(\beta_{jk},\gamma_{jk}=0|\eta_k=1) \right].\nonumber
\end{align}
\normalsize
where the last equation is because of the assumption $q(\beta_{jk},\gamma_{jk}|\eta_k)=q(\beta_{jk}|\gamma_{jk},\eta_k)q(\gamma_{jk})$ and $\bfgamma_{-jk}$ is a vector obtained by removing the $jk$-th term in $\bfgamma$. $\mathbb{E}_{j'\neq j|k}(\cdot)$ denotes taking the expectation with respect to all variables inside the $k$-th group except the $j$-th one. Again, given $q(\gamma_{jk})$, when $\gamma_{jk}=1$, we can further derive with a similar procedure from the expectation in Equation (6) that:
\small
\begin{eqnarray}
  \begin{aligned}
  &\mathbb{E}_{q(\beta_{jk}|\eta_k=1,\gamma_{jk}=1)}\left[\mathbb{E}_{j'\neq j|k}\left[ \mathbb{E}_{k'\neq k}\log\Pr(\mathbf{y},\bm{\eta}_{-k},\eta_k=1,\bm{\gamma}_{-jk},\gamma_{jk}=1,\bm{\beta})\right] - \log q(\beta_{jk},\gamma_{jk}=1|\eta_k=1)\right]\\
  =&\mathbb{E}_{q(\beta_{jk}|\eta_k=1,\gamma_{jk}=1)}\left[\mathbb{E}_{j'\neq j|k}\left[ \mathbb{E}_{k'\neq k}\log\Pr(\mathbf{y},\bm{\eta}_{-k},\eta_k=1,\bm{\gamma}_{-jk},\gamma_{jk}=1,\bm{\beta})\right] - \log q(\beta_{jk}|\eta_k=1,\gamma_{jk}=1) \right] + \mathrm{const},
  \end{aligned}
\end{eqnarray}
\normalsize
which is a KL Divergence between $\mathbb{E}_{j'\neq j|k}\left[ \mathbb{E}_{k'\neq k}\log\Pr(\mathbf{y},\bm{\eta}_{-k},\eta_k=1,\bm{\gamma}_{-jk},\gamma_{jk}=1,\bm{\beta})\right]$ and $q(\beta_{jk}|\eta_k=1,\gamma_{jk}=1)$ given $\eta_k=1$ and $\gamma_{jk}=1$. Hence the optimal form of $q^*(\beta_{jk}|\eta_k=1,\gamma_{jk}=1)$ is given by
\begin{equation}
  \log q^*(\beta_{jk}|\eta_k=1,\gamma_{jk}=1)=\mathbb{E}_{j'\neq j|k}\left[ \mathbb{E}_{k'\neq k}\log\Pr(\mathbf{y},\bm{\eta}_{-k},\eta_k=1,\bm{\gamma}_{-jk},\gamma_{jk}=1,\bm{\beta})\right].
\end{equation}
Here we only derive the case when $\eta_k=\gamma_{jk}=1$, other cases can be easily derived following the same procedure. Since both $\eta_k$ and $\gamma_{jk}$ are from Bernoulli distribution, with the expression in equation (8), we can first impose some variational parameters on $q(\gamma_{jk})$ and $q(\eta_k)$, then derive the conditional distribution of $\beta_{jk}$ given $\eta_k$ and $\gamma_{jk}$, and lastly optimize the lower bound to find the variational parameters.\\
First, we derive $q(\beta_{jk}|\eta_k,\gamma_{jk})$ which involves the joint probability function. The logarithm of the joint probability function is given as
\begin{eqnarray}
  \begin{aligned}
  \mathrm{log}\mathrm{Pr}(\mathbf{y},\bm{\eta},\bm{\gamma},\bm{\beta}|\mathbf{X},\mathbf{Z};\bm{\theta}) = &-\frac{n}{2}\mathrm{log}(2\pi\sigma_{e}^{2})-\frac{\mathbf{y}^{T}\mathbf{y}}{2\sigma_{e}^{2}}-\frac{(\mathbf{Z}\bm{\omega})^{T}(\mathbf{Z}\bm{\omega})}{2\sigma_{e}^{2}} \\
  &+\frac{\sum_{k}^{K}\sum_{j}^{l_{k}}\eta_{k}\gamma_{jk}\beta_{jk}\mathbf{x}_{jk}^{T}\mathbf{y}}{\sigma_{e}^{2}}+\frac{\mathbf{y}^{T}(\mathbf{Z}\bm{\omega})}{\sigma_{e}^{2}}-\frac{\sum_{k}^{K}\sum_{j}^{l_{k}}\eta_{k}\gamma_{jk}\beta_{jk}\mathbf{x}_{jk}^{T}(\mathbf{Z}\bm{\omega})}{\sigma_{e}^{2}}\\
  &-\frac{1}{2\sigma_{e}^{2}}\sum_{k}^{K}\sum_{j}^{l_{k}}\left(\left(\eta_{k}\gamma_{jk}\beta_{jk}\right)^{2}\mathbf{x}_{jk}^{T}\mathbf{x}_{jk}\right)\\
  &-\frac{1}{2\sigma_{e}^{2}}\left(\sum_{k}^{K}\sum_{k^{\prime}\neq k}^{K}\sum_{j}^{j_{k}}\sum_{j^{\prime}}^{l_{k^{\prime}}}\left(\eta_{k^{\prime}}\gamma_{j^{\prime} k^{\prime}}\beta_{j^{\prime} k^{\prime}}\right)\left(\eta_{k}\gamma_{jk}\beta_{jk}\right)\mathbf{x}_{j^{\prime} k^{\prime}}^{T}\mathbf{x}_{jk}\right) \\
  &-\frac{1}{2\sigma_{e}^{2}}\left(\sum_{k}^{K}\sum_{j}^{l_{k}}\sum_{j^{\prime}\neq j}^{l_{k}}\left(\eta_{k}\gamma_{j^{\prime} k}\beta_{j^{\prime} k}\right)\left(\eta_{k}\gamma_{jk}\beta_{jk}\right)\mathbf{x}_{j^{\prime} k}^{T}\mathbf{x}_{jk}\right)  \\
  &-\frac{p}{2}\mathrm{log}(2\pi\sigma_{\beta}^{2})-\frac{1}{2\sigma_{\beta}^{2}}\sum_{k=1}^{K}\sum_{j=1}^{l_{k}}\beta_{jk}^{2}  \\
  &+\mathrm{log}(\alpha)\sum_{k=1}^{K}\sum_{j=1}^{l_{k}}\gamma_{jk}+\mathrm{log}(1-\alpha)\sum_{k=1}^{K}\sum_{j=1}^{l_{k}}(1-\gamma_{jk})  \\
  &+\mathrm{log}(\pi)\sum_{k=1}^{K}\eta_{k}+\mathrm{log}(1-\pi)\sum_{k=1}^{K}(1-\eta_{k}).
  \end{aligned}
\end{eqnarray}
To find the optimal form in Equation (8), We then rearrange Equation (9) and only retain the terms regarding $jk$
\begin{eqnarray}
  \begin{aligned}
  &\mathrm{log}\mathrm{Pr}(\mathbf{y},\bm{\eta},\bm{\gamma},\bm{\beta}|\mathbf{X},\mathbf{Z};\bm{\theta}) \\
  =&-\frac{n}{2}\mathrm{log}(2\pi\sigma_{e}^{2})-\frac{\mathbf{y}^{T}\mathbf{y}}{2\sigma_{e}^{2}} \\
  &+\frac{\eta_{k}\gamma_{jk}\beta_{jk}\mathbf{x}_{jk}^{T}\mathbf{y}}{\sigma_{e}^{2}}-\frac{\eta_{k}\gamma_{jk}\beta_{jk}\mathbf{x}_{jk}^{T}(\mathbf{Z}\bm{\omega})}{\sigma_{e}^{2}}\\
  &-\frac{1}{2\sigma_{e}^{2}}\left(\left(\eta_{k}\gamma_{jk}\beta_{jk}\right)^{2}\mathbf{x}_{jk}^{T}\mathbf{x}_{jk}\right)\\
  &-\frac{1}{2\sigma_{e}^{2}}\left(\sum_{k'\neq k}^{K}\sum_{j'}^{l_{k'}}\left(\eta_{k}\gamma_{jk}\beta_{jk}\right)\left(\eta_{k}\gamma_{j'k'}\beta_{j'k'}\right)\mathbf{x}_{jk}^{T}\mathbf{x}_{j'k'}\right)\\
  &-\frac{1}{2\sigma_{e}^{2}}\left(\sum_{j'\neq j}^{l_{k}}\left(\eta_{k}\gamma_{jk}\beta_{jk}\right)\left(\eta_{k}\gamma_{j'k}\beta_{j'k}\right)\mathbf{x}_{jk}^{T}\mathbf{x}_{j'k}\right)-\frac{1}{2\sigma_{\beta}^{2}}\beta_{jk}^{2}\\
  &+\mathrm{log}(\alpha)\gamma_{jk}+\mathrm{log}(1-\alpha)(1-\gamma_{jk})\\
  &+\mathrm{log}(\pi)\eta_{k}+\mathrm{log}(1-\pi)(1-\eta_{jk})\\
  &+\mathrm{const}.
  \end{aligned}
\end{eqnarray}
Now we can derive the $\log\ q(\beta_{jk}|\eta_{k}=1,\gamma_{jk}=1)$ by taking the expectation in Equation (8). When $\eta_{k}=\gamma_{jk}=1\Leftrightarrow\eta_{k}\gamma_{jk}=1$, we have
\begin{eqnarray}
  \begin{aligned}
  &\log\ q(\beta_{jk}|\eta_{k}=1,\gamma_{jk}=1)\\
  =&\left(-\frac{1}{2\sigma_{e}^{2}}\mathbf{x}_{jk}^{T}\mathbf{x}_{jk}-\frac{1}{2\sigma_{\beta}^{2}}\right)\beta_{jk}^{2}\\
  &+\left(\frac{\mathbf{x}_{jk}^{T}(\mathbf{y}-\mathbf{Z}\bm{\omega})-\sum_{k'\neq k}^{K}\sum_{j'}^{l_{k'}}\mathbb{E}_{k'\neq k}[\eta_{k'}\gamma_{j'k'}\beta_{j'k'}]\mathbf{x}_{jk}^{T}\mathbf{x}_{j'k'}-\sum_{j'\neq j}^{l_{k}}\mathbb{E}_{j'\neq j|k}[\gamma_{j'k}\beta_{j'k}]\mathbf{x}_{jk}^{T}\mathbf{x}_{j'k}}{\sigma_{e}^{2}}\right)\beta_{jk}\\
  &+\mathrm{const}.
  \end{aligned}
\end{eqnarray}
Since Equation (11) is a quadratic form of $\beta_{jk}$, the posterior of $q(\beta_{jk}|\eta_{k}=1,\gamma_{jk}=1)$ follows a Gaussian of the form $\mathcal{N}(\mu_{jk},s_{jk}^{2})$, where
\begin{eqnarray}
  \begin{aligned}
  s_{jk}^{2}&=\frac{\sigma_{e}^{2}}{\mathbf{x}_{jk}^{T}\mathbf{x}_{jk}+\frac{\sigma_{e}^{2}}{\sigma_{\beta}^{2}}}\\
  \mu_{jk}&=\frac{\mathbf{x}_{jk}^{T}(\mathbf{y}-\mathbf{Z}\bm{\omega})-\sum_{k'\neq k}^{K}\sum_{j'}^{l_{k'}}\mathbb{E}_{k'\neq k}[\eta_{k'}\gamma_{j'k'}\beta_{j'k'}]\mathbf{x}_{jk}^{T}\mathbf{x}_{j'k'}-\sum_{j'\neq j}^{l_{k}}\mathbb{E}_{j'\neq j|k}[\gamma_{j'k}\beta_{j'k}]\mathbf{x}_{jk}^{T}\mathbf{x}_{j'k}}{\mathbf{x}_{jk}^{T}\mathbf{x}_{jk}+\frac{\sigma_{e}^{2}}{\sigma_{\beta}^{2}}}.
  \end{aligned}
\end{eqnarray}
Similarly, for $\eta_{k}\gamma_{jk}=0$, we have
\begin{equation}
\mathrm{log}\ q(\beta_{jk}|\eta_{k}\gamma_{jk}=0)=-\frac{1}{2\sigma_{\beta}^{2}}\beta_{jk}^{2}+\mathrm{const},
\end{equation}
which implies that $q(\beta_{jk}|\eta_{k}\gamma_{jk}=0)\sim \mathcal{N}(0,\sigma_{\beta}^{2})$. Thus, the conditional posterior of $\beta_{jk}$ is exactly the same as the prior if this variable is irrelevant in either one of the two levels ($\eta_{k}\gamma_{jk}=0$). Now we turn to $q(\eta_k)$ and $q(\gamma_{jk})$. Denote $\pi_k=q(\eta_k)$ and $\alpha_{jk}=q(\gamma_{jk})$, we have
\begin{equation}
q(\bm{\eta},\bm{\gamma},\bm{\beta})=\prod_{k}^{K}\left(\pi_{k}^{\eta_{k}}(1-\pi_{k})^{1-\eta_{k}}\prod_{j}^{l_{k}}\left(\alpha_{jk}^{\gamma_{jk}}(1-\alpha_{jk})^{1-\gamma_{jk}}\mathcal{N}(\mu_{jk},s_{jk}^{2})^{\eta_{k}\gamma_{jk}}\mathcal{N}(0,\sigma_{\beta})^{1-\eta_{k}\gamma_{jk}}\right)\right).
\end{equation}
And the second term in $\mathcal{L}(q)$ can be derived as:
\begin{eqnarray}
  \begin{aligned}
  &-\mathbb{E}_q[\mathrm{log}q(\bm{\eta},\bm{\gamma},\bm{\beta})]\\
  =&-\mathbb{E}_{q}\left[\sum_{k}^{K}\left(\eta_{k}\mathrm{log}(\pi_{k})+(1-\eta_{k})\mathrm{log}(1-\pi_{k})\right)\right]\\
  &-\mathbb{E}_{q}\left[\sum_{k}^{K}\sum_{j}^{l_{k}}\eta_{k}\gamma_{jk}\mathrm{log}\mathcal{N}(\mu_{jk},s_{jk}^{2})+(1-\eta_{k}\gamma_{jk})\mathcal{N}(0,\sigma_{\beta}^{2})+\gamma_{jk}\mathrm{log}(\alpha_{jk})+(1-\gamma_{jk})\mathrm{log}(1-\alpha_{jk})\right]\\
  =&-\sum_{k}^{K}\mathbb{E}_{q}\left[\left(\eta_{k}\mathrm{log}(\pi_{k})+(1-\eta_{k})\mathrm{log}(1-\pi_{k})\right)\right]\\
  &-\sum_{k}^{K}\sum_{j}^{l_{k}}\mathbb{E}_{q}\left[\eta_{k}\gamma_{jk}\mathrm{log}\mathcal{N}(\mu_{jk},s_{jk}^{2})+(1-\eta_{k}\gamma_{jk})\mathcal{N}(0,\sigma_{\beta}^{2})+\gamma_{jk}\mathrm{log}(\alpha_{jk})+(1-\gamma_{jk})\mathrm{log}(1-\alpha_{jk})\right]\\
  =&-\sum_{k}^{K}\sum_{j}^{l_{k}}\mathbb{E}_{\eta_{k},\gamma_{jk}}\{\mathbb{E}_{\beta|\eta_{k}=1,\gamma_{jk}=1}[\mathrm{log}\mathcal{N}(\mu_{jk},s_{jk}^{2})]+\mathbb{E}_{\beta|\eta_{k}=1,\gamma_{jk}=0}[\mathrm{log}\mathcal{N}(0,\sigma_{\beta}^{2})]\\
  &+\mathbb{E}_{\beta|\eta_{k}=0,\gamma_{jk}=1}[\mathrm{log}\mathcal{N}(0,\sigma_{\beta}^{2})]+\mathbb{E}_{\beta|\eta_{k}=0,\gamma_{jk}=0}[\mathrm{log}\mathcal{N}(0,\sigma_{\beta}^{2})]\}\\
  &-\sum_{k}^{K}\sum_{j}^{l_{k}}\mathbb{E}_{q}[\gamma_{jk}\mathrm{log}(\alpha_{jk})+(1-\gamma_{jk})\mathrm{log}(1-\alpha_{jk})]-\sum_{k}^{K}\mathbb{E}_{q}[\eta_{k}\mathrm{log}(\pi_{k})+(1-\eta_{k})\mathrm{log}(1-\pi_{k})]\\
  \end{aligned}
\end{eqnarray}
Note that $-\mathbb{E}_{\beta|\eta_{k}=1,\gamma_{jk}=1}[\mathrm{log}\mathcal{N}(\mu_{jk},s_{jk}^{2})]$ is the entropy of Gaussian, so we have\\ $-\mathbb{E}_{\beta|\eta_{k}=1,\gamma_{jk}=1}[\mathrm{log}\mathcal{N}(\mu_{jk},s_{jk}^{2})]=\frac{1}{2}\mathrm{log}(s_{jk}^{2})+\frac{1}{2}(1+\mathrm{log}(2\pi))$, similarly, $-\mathbb{E}_{\beta|\eta_{k}=1,\gamma_{jk}=0}[\mathrm{log}\mathcal{N}(\mu_{jk},s_{jk}^{2})]=\frac{1}{2}\mathrm{log}(\sigma_{\beta}^{2})+\frac{1}{2}(1+\mathrm{log}(2\pi))$ and so on. Consequently, Equation (15) can be further derived in:

\begin{align}
  &-\mathbb{E}[\mathrm{log}q(\bm{\eta},\bm{\gamma},\bm{\beta})]\nonumber\\
  =&\sum_{k}^{K}\sum_{j}^{l_{k}}\{[\frac{1}{2}\mathrm{log}(s_{jk}^{2})+\frac{1}{2}(1+\mathrm{log}(2\pi))]\pi_{k}\alpha_{jk}+[\frac{1}{2}\mathrm{log}(\sigma_{\beta}^{2})+\frac{1}{2}(1+\mathrm{log}(2\pi))](1-\pi_{k}\alpha_{jk})\nonumber\\
  &-\alpha_{jk}\mathrm{log}(\alpha_{jk})-(1-\alpha_{jk})\mathrm{log}(1-\alpha_{jk})\}-\sum_{k}^{K}\{\pi_{k}\mathrm{log}(\pi_{k})+(1-\pi_{k})\mathrm{log}(1-\pi_{k})\}\nonumber\\
  =&\sum_{k}^{K}\sum_{j}^{l_{k}}\frac{1}{2}\pi_{k}\alpha_{jk}(\mathrm{log}s_{jk}^{2}-\mathrm{log}\sigma_{\beta}^{2})+\frac{p}{2}\mathrm{log}(\sigma_{\beta}^{2})+\frac{p}{2}+\frac{p}{2}\mathrm{log}(2\pi)\\
  &-\sum_{k}^{K}\sum_{j}^{l_{k}}[\alpha_{jk}\mathrm{log}(\alpha_{jk})+(1-\alpha_{jk})\mathrm{log}(1-\alpha_{jk})]-\sum_{k}^{K}[\pi_{k}\mathrm{log}(\pi_{k})+(1-\pi_{k})\mathrm{log}(1-\pi_{k})].\nonumber
  \end{align}

Combine (16) with (9), the lower bound is obtained as follow:
\begin{eqnarray}
  \begin{aligned}
  &\mathbb{E}_q[\mathrm{log}\mathrm{Pr}(\mathbf{y},\bm{\eta},\bm{\gamma},\bm{\beta}|\mathbf{X};\bm{\theta})]-\mathbb{E}_q[\mathrm{log}q(\bm{\eta},\bm{\gamma},\bm{\beta})] \\
  =&-\frac{n}{2}\mathrm{log}(2\pi\sigma_{e}^{2})-\frac{\mathbf{y}^{T}\mathbf{y}}{2\sigma_{e}^{2}}-\frac{(\mathbf{Z}\bm{\omega})^{T}(\mathbf{Z}\bm{\omega})}{2\sigma_{e}^{2}}\\
  &+\frac{\sum_{k}^{K}\sum_{j}^{l_{k}}\mathbb{E}_q\left[\eta_{k}\gamma_{jk}\beta_{jk}\right]\mathbf{x}_{jk}^{T}\mathbf{y}}{\sigma_{e}^{2}}+\frac{\mathbf{y}^{T}(\mathbf{Z}\bm{\omega})}{\sigma_{e}^{2}}-\frac{\sum_{k}^{K}\sum_{j}^{l_{k}}\mathbb{E}_q\left[\eta_{k}\gamma_{jk}\beta_{jk}\right]\mathbf{x}_{jk}^{T}(\mathbf{Z}\bm{\omega})}{\sigma_{e}^{2}}\\
  &-\frac{1}{2\sigma_{e}^{2}}\sum_{k}^{K}\sum_{j}^{l_{k}}\left(\mathbb{E}_q\left[\left(\eta_{k}\gamma_{jk}\beta_{jk}\right)^{2}\right]\mathbf{x}_{jk}^{T}\mathbf{x}_{jk}\right)\\
  &-\frac{1}{2\sigma_{e}^{2}}\left(\sum_{k}^{K}\sum_{k^{\prime}\neq k}^{K}\sum_{j}^{j_{k}}\sum_{j^{\prime}}^{l_{k^{\prime}}}\mathbb{E}_q\left[\eta_{k^{\prime}}\gamma_{j^{\prime} k^{\prime}}\beta_{j^{\prime} k^{\prime}}\right]\mathbb{E}_q\left[\eta_{k}\gamma_{jk}\beta_{jk}\right]\mathbf{x}_{j^{\prime} k^{\prime}}^{T}\mathbf{x}_{jk}\right) \\
  &-\frac{1}{2\sigma_{e}^{2}}\left(\sum_{k}^{K}\sum_{j}^{l_{k}}\sum_{j^{\prime}\neq j}^{l_{k}}\mathbb{E}_q\left[\eta_{k}^{2}\gamma_{j^{\prime} k}\beta_{j^{\prime} k}\gamma_{jk}\beta_{jk}\right]\mathbf{x}_{j^{\prime} k}^{T}\mathbf{x}_{jk}\right)  \\
  &-\frac{p}{2}\mathrm{log}(2\pi\sigma_{\beta}^{2})-\frac{1}{2\sigma_{\beta}^{2}}\sum_{k=1}^{K}\sum_{j=1}^{l_{k}}\mathbb{E}_q\left[\beta_{jk}^{2}\right]  \\
  &+\mathrm{log}(\alpha)\sum_{k=1}^{K}\sum_{j=1}^{l_{k}}\mathbb{E}_q\left[\gamma_{jk}\right]+\mathrm{log}(1-\alpha)\sum_{k=1}^{K}\sum_{j=1}^{l_{k}}\mathbb{E}_q\left[1-\gamma_{jk}\right]  \\
  &+\mathrm{log}(\pi)\sum_{k=1}^{K}\mathbb{E}_q\left[\eta_{k}\right]+\mathrm{log}(1-\pi)\sum_{k=1}^{K}\mathbb{E}_q\left[1-\eta_{k}\right]\\
  &+\sum_{k}^{K}\sum_{j}^{l_{k}}\frac{1}{2}\pi_{k}\alpha_{jk}(\mathrm{log}s_{jk}^{2}-\mathrm{log}\sigma_{\beta}^{2})+\frac{p}{2}\mathrm{log}(\sigma_{\beta}^{2})+\frac{p}{2}+\frac{p}{2}\mathrm{log}(2\pi)\\
  &-\sum_{k}^{K}\sum_{j}^{l_{k}}[\alpha_{jk}\mathrm{log}(\alpha_{jk})]-\sum_{k}^{K}\sum_{j}^{l_{k}}[(1-\alpha_{jk})\mathrm{log}(1-\alpha_{jk})]\\
  &-\sum_{k}^{K}[\pi_{k}\mathrm{log}(\pi_{k})]-\sum_{k}^{K}[(1-\pi_{k})\mathrm{log}(1-\pi_{k})],
  \end{aligned}
\end{eqnarray}
where expectations in are derived as follows: 
\begin{equation}
  \mathbb{E}_q\left[\eta_{k}\right]=\pi_{k},\ \ \ \ \mathbb{E}_q\left[\gamma_{jk}\right]=\alpha_{jk},
\end{equation}
\begin{align}
  \mathbb{E}\left[\eta_{k}\gamma_{jk}\beta_{jk}\right]&=\sum_{\gamma_{jk}}\sum_{\eta_k}\int_{\beta_{jk}}\eta_{k}\gamma_{jk}\beta_{jk}q(\beta_{jk}|\eta_k,\gamma_{jk})q(\eta_k)q(\gamma_{jk})d\beta_{jk}\nonumber\\
  &=\pi_k\alpha_{jk}\cdot\mu_{jk} + (1-\pi_k\alpha_{jk})\cdot0\nonumber\\
  &=\pi_k\alpha_{jk}\mu_{jk}
\end{align}
\begin{align}
  \mathbb{E}_q\left[\beta_{jk}^{2}\right]&=\int_{\beta_{jk}}\beta_{jk}^2q(\beta_{jk})d\beta_{jk}\nonumber\\
  &=\sum_{\eta_k}\sum_{\gamma_{jk}}\int_{\beta_{jk}}\beta_{jk}^2q(\beta_{jk}|\eta_k,\gamma_{jk})q(\eta_k)q(\gamma_{jk})d\beta_{jk}\nonumber\\
  &=\int_{\beta_{jk}}\beta_{jk}^2\cdot\left[ \pi_k\alpha_{jk}\mathcal{N}(\mu_{jk},s_{jk}^2) + (1-\pi_k\alpha_{jk})\mathcal{N}(0,\sigma_{\beta}^2) \right]d\beta_{jk}\nonumber\\
  &=\pi_{k}\alpha_{jk}(s_{jk}^{2}+\mu_{jk}^{2})+(1-\pi_{k}\alpha_{jk})\sigma_{\beta}^{2}
\end{align}
\begin{align}
  \mathbb{E}_q\left[(\eta_{k}\gamma_{jk}\beta_{jk})^{2}\right]&=\sum_{\eta_k}\sum_{\gamma_{jk}}\int_{\beta_{jk}} \eta_{k}\gamma_{jk}\beta_{jk}^{2}q(\beta_{jk}|\eta_k,\gamma_{jk})q(\eta_k)q(\gamma_{jk}) d\beta_{jk}\nonumber\\
  &=\pi_k\alpha_{jk}\int_{\beta_{jk}}\beta_{jk}^{2}\mathcal{N}(\mu_{jk},s_{jk}^2))d\beta_{jk}\nonumber\\
  &=\pi_{k}\alpha_{jk}(s_{jk}^{2}+\mu_{jk}^{2})
\end{align}
\begin{align}
  &\mathbb{E}_q\left[\eta_{k}^{2}\gamma_{j^{\prime} k}\beta_{j^{\prime} k}\gamma_{jk}\beta_{jk}\right]\nonumber\\
  =&\mathbb{E}_q\left[\eta_{k}\gamma_{j^{\prime} k}\beta_{j^{\prime} k}\gamma_{jk}\beta_{jk}\right]\nonumber\\
  =&\sum_{\eta_k}\sum_{\gamma_{jk},\gamma_{j'k}}\int\int\eta_{k}\gamma_{j^{\prime} k}\beta_{j^{\prime} k}\gamma_{jk}\beta_{jk}q(\beta_{jk}|\eta_k,\gamma_{jk})q(\gamma_{jk})q(\beta_{j'k}|\eta_k,\gamma_{j'k})q(\gamma_{j'k})q(\eta_k)d\beta_{jk}d\beta_{j'k}\nonumber\\
  =&\pi_{k}\alpha_{j^{\prime}k}\mu_{j^{\prime}k}\alpha_{jk}\mu_{jk}
\end{align}
We plug in the evaluations from Equation (18) to (22), $\mathcal{L}(q)$ in Equation (17) then becomes
\begin{eqnarray}
  \begin{aligned}
  &\mathbb{E}_q[\mathrm{log}\mathrm{Pr}(\mathbf{y},\bm{\eta},\bm{\gamma},\bm{\beta}|\mathbf{X};\bm{\theta})]-\mathbb{E}_q[\mathrm{log}q(\bm{\eta},\bm{\gamma},\bm{\beta})] \\
  =&-\frac{n}{2}\mathrm{log}(2\pi\sigma_{e}^{2})-\frac{||\mathbf{y}-\mathbf{Z}\bm{\omega}-\sum_{k}^{K}\sum_{j}^{l_{k}}\pi_{k}\alpha_{jk}\mu_{jk}\mathbf{x}_{jk}||^{2}}{2\sigma_{e}^{2}} \\
  &-\frac{1}{2\sigma_{e}^{2}}\sum_{k}^{K}\sum_{j}^{l_{k}}\underbrace{[\pi_{k}\alpha_{jk}(s_{jk}^{2}+\mu_{jk}^{2})-(\pi_{k}\alpha_{jk}\mu_{jk})^{2}]}_{\text{Var}[\eta_{k}\gamma_{jk}\beta_{jk}]}\mathbf{x}_{jk}^{T}\mathbf{x}_{jk}\\
  &-\frac{1}{2\sigma_{e}^{2}}\sum_{k}^{K}\left(\pi_{k}-\pi_{k}^{2}\right)\left(\sum_{j}^{l_{k}}\sum_{j^{\prime}\neq j}^{l_{k}}\alpha_{j^{\prime}k}\mu_{j^{\prime}k}\alpha_{jk}\mu_{jk}\mathbf{x}_{j^{\prime}k}^{T}\mathbf{x}_{jk}\right)\\
  &-\frac{p}{2}\mathrm{log}(2\pi\sigma_{\beta}^{2})-\frac{1}{2\sigma_{\beta}^{2}}\sum_{k=1}^{K}\sum_{j=1}^{l_{k}}[\pi_{k}\alpha_{jk}(s_{jk}^{2}+\mu_{jk}^{2})+(1-\pi_{k}\alpha_{jk})\sigma_{\beta}^{2}]\\
  &+\sum_{k=1}^{K}\sum_{j=1}^{l_{k}}\alpha_{jk}\mathrm{log}(\frac{\alpha}{\alpha_{jk}})+\sum_{k=1}^{K}\sum_{j=1}^{l_{k}}(1-\alpha_{jk})\mathrm{log}(\frac{1-\alpha}{1-\alpha_{jk}})\\
  &+\sum_{k=1}^{K}\pi_{k}\mathrm{log}(\frac{\pi}{\pi_{k}})+\sum_{k=1}^{K}(1-\pi_{k})\mathrm{log}(\frac{1-\pi}{1-\pi_{k}})\\
  &+\sum_{k}^{K}\sum_{j}^{l_{k}}\frac{1}{2}\pi_{k}\alpha_{jk}\mathrm{log}(\frac{s_{jk}^{2}}{\sigma_{\beta}^{2}})+\frac{p}{2}\mathrm{log}(\sigma_{\beta}^{2})+\frac{p}{2}+\frac{p}{2}\mathrm{log}(2\pi)\\
  \end{aligned}
\end{eqnarray}
To get $\pi_{k}$ and $\alpha_{jk}$, we set $$\frac{\partial\mathbb{E}_q[\mathrm{log}\mathrm{Pr}(\mathbf{y},\bm{\eta},\bm{\gamma},\bm{\beta}|\mathbf{X};\bm{\theta})]-\mathbb{E}_q[\mathrm{log}q(\bm{\eta},\bm{\gamma},\bm{\beta})]}{\partial\pi_{k}}=0,$$ $$\frac{\partial\mathbb{E}_q[\mathrm{log}\mathrm{Pr}(\mathbf{y},\bm{\eta},\bm{\gamma},\bm{\beta}|\mathbf{X};\bm{\theta})]-\mathbb{E}_q[\mathrm{log}q(\bm{\eta},\bm{\gamma},\bm{\beta})]}{\partial\alpha_{jk}}=0,$$
which gives
\begin{eqnarray}
  \begin{aligned}
  &\pi_{k}=\frac{1}{1+\mathrm{exp}(-u_{k})},\\
  \mathrm{where}\ u_{k}=&\mathrm{log}\frac{\pi}{1-\pi}+\frac{1}{2}\sum_{j}^{l_{k}}\alpha_{jk}\left(\mathrm{log}\frac{s_{jk}^{2}}{\sigma_{\beta}^{2}}+\frac{\mu_{jk}^{2}}{s_{jk}^{2}}\right);\\
  \mathrm{and}\ &\alpha_{jk}=\frac{1}{1+\mathrm{exp}(-v_{jk})},\\
  \mathrm{where}\  v_{jk}=&\mathrm{log}\frac{\alpha}{1-\alpha}+\frac{1}{2}\pi_{k}\left(\mathrm{log}\frac{s_{jk}^{2}}{\sigma_{\beta}^{2}}+\frac{\mu_{jk}^{2}}{s_{jk}^{2}}\right).
  \end{aligned}
\end{eqnarray}
The derivation is as follow:
\begin{eqnarray}
  \begin{aligned}
   u_{k}=&\mathrm{log}\frac{\pi}{1-\pi}+\frac{1}{2}\sum_{j}^{l_{k}}\alpha_{jk}\mathrm{log}\frac{s_{jk}^{2}}{\sigma_{\beta}^{2}}\\
   &+\frac{\sum_{j}^{l_{k}}\alpha_{jk}\mu_{jk}\mathbf{x}_{jk}^{T}\mathbf{y}}{\sigma_{e}^{2}}-\frac{\sum_{j}^{l_{k}}\alpha_{jk}\mu_{jk}\mathbf{x}_{jk}^{T}(\mathbf{Z}\bm{\omega})}{\sigma_{e}^{2}}-\frac{1}{2\sigma_{e}^{2}}\sum_{j}^{l_{k}}\alpha_{jk}(s_{jk}^{2}+\mu_{jk}^{2})\mathbf{x}_{jk}^{T}\mathbf{x}_{jk}\\
   &-\frac{1}{\sigma_{e}^{2}}\left(\sum_{k^{\prime}\neq k}^{K}\sum_{j}^{l_{k}}\sum_{j^{\prime}}^{l_{k^{\prime}}}\pi_{k^{\prime}}\alpha_{j^{\prime}k^{\prime}}\mu_{j^{\prime}k^{\prime}}\alpha_{jk}\mu_{jk}\mathbf{x}_{j^{\prime}k^{\prime}}^{T}\mathbf{x}_{jk}\right)\\
   &-\frac{1}{\sigma_{e}^{2}}\left(\sum_{j}^{l_{k}}\sum_{j^{\prime}\neq j}^{l_{k}}\alpha_{j^{\prime}k}\mu_{j^{\prime}k}\alpha_{jk}\mu_{jk}\mathbf{x}_{j^{\prime}k}^{T}\mathbf{x}_{jk}\right)\\
   &-\frac{1}{2\sigma_{\beta}}\sum_{j}^{l_{k}}\alpha_{jk}(s_{jk}^{2}+\mu_{jk}^{2})+\frac{1}{2}\sum_{j}^{l_{k}}\alpha_{jk}\\
   =&\mathrm{log}\frac{\pi}{1-\pi}+\frac{1}{2}\sum_{j}^{l_{k}}\alpha_{jk}\mathrm{log}\frac{s_{jk}^{2}}{\sigma_{\beta}^{2}}\\
   &+\alpha_{jk}\mu_{jk}\sum_{j}^{K}\underbrace{\left(\frac{\mathbf{x}_{jk}^{T}(\mathbf{y}-\mathbf{Z}\bm{\omega})-\sum_{k^{\prime}\neq k}^{K}\pi_{k^{\prime}}\sum_{j^{\prime}}^{l_{k^{\prime}}}\alpha_{j^{\prime}k^{\prime}}\mu_{j^{\prime}k^{\prime}}\mathbf{x}_{j^{\prime}k^{\prime}}^{T}\mathbf{x}_{jk}-\sum_{j^{\prime}\neq j}^{l_{k}}\alpha_{j^{\prime}k}\mu_{j^{\prime}k}\mathbf{x}_{j^{\prime}k}^{T}\mathbf{x}_{jk}}{\sigma_{e}^{2}}\right)}_{\mu_{jk}/s_{jk}^{2}}\\
   &-\frac{1}{2}\sum_{j}^{l_{k}}\alpha_{jk}(s_{jk}^{2}+\mu_{jk}^{2})\underbrace{\left(\frac{\mathbf{x}_{jk}^{T}\mathbf{x}_{jk}}{\sigma_{e}^{2}}+\frac{1}{\sigma_{\beta}^{2}}\right)}_{1/s_{jk}^{2}}+\frac{1}{2}\sum_{j}^{l_{k}}\alpha_{jk}\\
   =&\mathrm{log}\frac{\pi}{1-\pi}+\frac{1}{2}\sum_{j}^{l_{k}}\alpha_{jk}\mathrm{log}\frac{s_{jk}^{2}}{\sigma_{\beta}^{2}}+\sum_{j}^{l_{k}}\frac{\alpha_{jk}\mu_{jk}^{2}}{s_{jk}^{2}}-\sum_{j}^{l_{k}}\frac{\alpha_{jk}\mu_{jk}^{2}}{2s_{jk}^{2}}\\
   =&\mathrm{log}\frac{\pi}{1-\pi}+\frac{1}{2}\sum_{j}^{l_{k}}\alpha_{jk}\left(\mathrm{log}\frac{s_{jk}^{2}}{\sigma_{\beta}^{2}}+\frac{\mu_{jk}^{2}}{s_{jk}^{2}}\right),\\
  \end{aligned}
\end{eqnarray}
where we have used Equation (11) in the third equation. Derivation of $v_{jk}$ follows the same procedure.
\subsubsection{M-step}
At M-step, we update the parameters $\bm{\theta}=\{\alpha,\pi,\sigma_{\beta}^{2},\sigma_{e}^{2},\bm{\omega}\}$. By setting 
$\frac{\partial\mathbb{E}_q[\mathrm{log}\mathrm{Pr}(\mathbf{y},\bm{\eta},\bm{\gamma},\bm{\beta}|\mathbf{X},\mathbf{Z};\bm{\theta})]}{\partial\sigma_{e}^{2}}=0,$
we have
\begin{equation}
  \begin{aligned}
  \sigma_{e}^{2}=&\frac{||\mathbf{y}-\mathbf{Z}\bm{\omega}-\sum_{k}^{K}\sum_{j}^{l_{k}}\pi_{k}\alpha_{jk}\mu_{jk}\mathbf{x}_{jk}||^{2}}{n}\\
  &+\frac{\sum_{k}^{K}\sum_{j}^{l_{k}}[\pi_{k}\alpha_{jk}(s_{jk}^{2}+\mu_{jk}^{2})-(\pi_{k}\alpha_{jk}\mu_{jk})^{2}]\mathbf{x}_{jk}^{T}\mathbf{x}_{jk}}{n}\\
  &+\frac{\sum_{k}^{K}(\pi_{k}-\pi_{k}^{2})[\sum_{j}^{l_{k}}\sum_{j^{\prime}}^{l_{k}}\alpha_{j^{\prime}k}\mu_{j^{\prime}k}\alpha_{jk}\mu_{jk}]\mathbf{x}_{j^{\prime}k}^{T}\mathbf{x}_{jk}}{n}.
  \end{aligned}
\end{equation}
To get $\sigma_{\beta}^{2}$, we set
$\frac{\partial \mathcal{L}(q)}{\partial\sigma_{\beta}^{2}}=0,$
which gives
\begin{equation}
\sigma_{\beta}^{2}=\frac{\sum_{k}^{K}\sum_{j}^{l_{k}}\pi_{k}\alpha_{jk}(s_{jk}^{2}+\mu_{jk}^{2})}{\sum_{k}^{K}\sum_{j}^{l_{k}}\pi_{k}\alpha_{jk}}.
\end{equation}
Accordingly, 
\begin{equation}
\alpha=\frac{1}{p}\sum_{k}^{K}\sum_{j}^{l_{k}}\alpha_{jk},
\end{equation}
\begin{equation}
\pi=\frac{1}{K}\sum_{k}^{K}\pi_{k}.
\end{equation}
\begin{equation}
\bm{\omega}=(\mathbf{Z}^{T}\mathbf{Z})^{-1}\mathbf{Z}^{T}(\mathbf{y}-\sum_{k}^{K}\sum_{j}^{l_{k}}\pi_{k}\alpha_{jk}\mu_{jk}\mathbf{x}_{jk}).
\end{equation}

\subsection{Variational EM Algorithm: Multi-task Learning with BIVAS}

\subsubsection{E-step}
Let $\bm{\theta}=\{\alpha,\pi,\sigma_{\beta_{j}}^{2},\sigma_{e_{j}}^{2},\bm{\omega}_{j}\}_{j=1}^{l}$ be the collection of model parameters. The joint probabilistic model is
\begin{eqnarray}
  \begin{aligned}
  &\mathrm{Pr}(\mathbf{y},\bm{\eta},\bm{\gamma},\bm{\beta}|\mathbf{X},\mathbf{Z};\theta) 
  =\mathrm{Pr}(\mathbf{y}|\bm{\eta},\bm{\gamma},\bm{\beta},\mathbf{X},\mathbf{Z},\theta)\mathrm{Pr}(\bm{\eta},\bm{\gamma},\bm{\beta}|\theta)  \\
   =&\prod_{j=1}^{L}\mathcal{N}(\mathbf{y}_{j}|\mathbf{Z}_{j}\bm{\omega}_{j}+\sum_{k}^{K}\eta_{k}\gamma_{jk}\beta_{jk}\mathbf{x}_{jk})\prod_{k=1}^{K}\pi^{\eta_{k}}(1-\pi)^{1-\eta_{k}}\prod_{j=1}^{L}\mathcal{N}(0,\sigma_{\beta_{j}}^{2})\alpha^{\gamma_{jk}}(1-\alpha)^{1-\gamma_{jk}}.
  \end{aligned}
\end{eqnarray}
The logarithm of the marginal likelihood is
\begin{eqnarray}
  \begin{aligned}
  \mathrm{log}\ p(\mathbf{y}|\mathbf{X},\mathbf{Z};\theta)&=\mathrm{log}\sum_{\bm{\gamma}}\sum_{\bm{\eta}}\int_{\bf{\beta}}\mathrm{Pr}(\mathbf{y},\bm{\eta},\bm{\gamma},\bm{\beta}|\mathbf{X},\mathbf{Z};\theta)d\bm{\beta}\\
  &\geq\sum_{\bm{\gamma}}\sum_{\bf{\eta}}\int_{\bf{\beta}}q(\bm{\eta},\bm{\gamma},\bm{\beta})\mathrm{log} \frac{\mathrm{Pr}(\mathbf{y},\bm{\eta},\bm{\gamma},\bm{\beta}|\mathbf{X},\mathbf{Z};\theta)}{q(\bm{\eta},\bm{\gamma},\bm{\beta})}\\
  &=\mathbb{E}_q[\mathrm{log}\mathrm{Pr}(\mathbf{y},\bm{\eta},\bm{\gamma},\bm{\beta}|\mathbf{X},\mathbf{Z};\theta)-q(\bm{\eta},\bm{\gamma},\bm{\beta})]\\
  &\equiv\mathcal{L}(q).
  \end{aligned}
\end{eqnarray}
Again, we assume that the variational distribution takes the form
\begin{equation}
q(\bm{\eta},\bm{\gamma},\bm{\beta})=\prod_{k}^{K}\left(q(\eta_{k})\prod_{j}^{L}(q(\beta_{jk}|\eta_{k},\gamma_{jk})q(\gamma_{jk}))\right).
\end{equation}
Actually, the variational approximation only assumes `between group' factorizability ($\prod_{k=1}^{K}q(\eta_{k},\gamma_{jk},\beta_{jk})$) because given the group, the tasks inside are independent due to model assumption. Follow the same procedure in Section 1.1, the optimal form of $q$ is given by
\begin{equation}
  \log q^*(\beta_{jk}|\eta_k=1,\gamma_{jk}=1)=\mathbb{E}_{j'\neq j|k}\left[ \mathbb{E}_{k'\neq k}\log\Pr(\mathbf{y},\bm{\eta}_{-k},\eta_k=1,\bm{\gamma}_{-jk},\gamma_{jk}=1,\bm{\beta})\right].
\end{equation}
The Equation (34) contains the logarithm of joint probability funcion, which is
\begin{eqnarray}
  \begin{aligned}
  \mathrm{log}\mathrm{Pr}(\mathbf{y},\bm{\eta},\bm{\gamma},\bm{\beta}|\mathbf{X},\mathbf{Z};\theta) = &\sum_{j=1}^{L}\Bigg\{-\frac{N_{j}}{2}\mathrm{log}(2\pi\sigma_{e_{j}}^{2})-\frac{\mathbf{y}_{j}^{T}\mathbf{y}_{j}}{2\sigma_{e_{j}}^{2}}-\frac{(\mathbf{Z}_{j}\bm{\omega}_{j})^{T}(\mathbf{Z}_{j}\bm{\omega}_{j})}{2\sigma_{e_{j}}^{2}} \\
  &+\frac{\sum_{k}^{K}\eta_{k}\gamma_{jk}\beta_{jk}\mathbf{x}_{jk}^{T}\mathbf{y}_{j}}{\sigma_{e_{j}}^{2}}+\frac{\mathbf{y}_{j}^{T}(\mathbf{Z}_{j}\bm{\omega}_{j})}{\sigma_{e_{j}}^{2}}-\frac{\sum_{k}^{K}\eta_{k}\gamma_{jk}\beta_{jk}\mathbf{x}_{jk}^{T}(\mathbf{Z}_{j}\bm{\omega}_{j})}{\sigma_{e_{j}}^{2}}\\
  &-\frac{1}{2\sigma_{e_{j}}^{2}}\sum_{k}^{K}\left(\left(\eta_{k}\gamma_{jk}\beta_{jk}\right)^{2}\mathbf{x}_{jk}^{T}\mathbf{x}_{jk}\right)\\
  &-\frac{1}{2\sigma_{e_{j}}^{2}}\left(\sum_{k}^{K}\sum_{k^{\prime}\neq k}^{K}\left(\eta_{k^{\prime}}\gamma_{j k^{\prime}}\beta_{jk^{\prime}}\right)\left(\eta_{k}\gamma_{jk}\beta_{jk}\right)\mathbf{x}_{j k^{\prime}}^{T}\mathbf{x}_{jk}\right)\Bigg\} \\
  &-\frac{K}{2}\sum_{j=1}^{L}\mathrm{log}(2\pi\sigma_{\beta_{j}}^{2})-\sum_{j=1}^{L}\frac{\sum_{k=1}^{K}\beta_{jk}^{2}}{2\sigma_{\beta_{j}}^{2}}  \\
  &+\mathrm{log}(\alpha)\sum_{k=1}^{K}\sum_{j=1}^{L}\gamma_{jk}+\mathrm{log}(1-\alpha)\sum_{k=1}^{K}\sum_{j=1}^{L}(1-\gamma_{jk})  \\
  &+\mathrm{log}(\pi)\sum_{k=1}^{K}\eta_{k}+\mathrm{log}(1-\pi)\sum_{k=1}^{K}(1-\eta_{k}).
  \end{aligned}
\end{eqnarray}

We then rearrange Equation (35) and retain the terms regarding $jk$
\begin{eqnarray}
\begin{aligned}
&\mathrm{log}\mathrm{Pr}(\mathbf{y},\bm{\eta},\bm{\gamma},\bm{\beta}|\mathbf{X},\mathbf{Z};\theta) \\
=&-\frac{N_{j}}{2}\mathrm{log}(2\pi\sigma_{e_{j}}^{2})-\frac{\mathbf{y}_{j}^{T}\mathbf{y}_{j}}{2\sigma_{e_{j}}^{2}} \\
&+\frac{\eta_{k}\gamma_{jk}\beta_{jk}\mathbf{x}_{jk}^{T}\mathbf{y}_{j}}{\sigma_{e_{j}}^{2}}-\frac{\eta_{k}\gamma_{jk}\beta_{jk}\mathbf{x}_{jk}^{T}(\mathbf{Z}_{j}\bm{\omega}_{j})}{\sigma_{e_{j}}^{2}}\\
&-\frac{1}{2\sigma_{e_{j}}^{2}}\left(\left(\eta_{k}\gamma_{jk}\beta_{jk}\right)^{2}\mathbf{x}_{jk}^{T}\mathbf{x}_{jk}\right)\\
&-\frac{1}{2\sigma_{e_{j}}^{2}}\left(\sum_{k'\neq k}^{K}\left(\eta_{k}\gamma_{jk}\beta_{jk}\right)\left(\eta_{k'}\gamma_{jk'}\beta_{jk'}\right)\mathbf{x}_{jk}^{T}\mathbf{x}_{jk'}\right)-\frac{1}{2\sigma_{\beta_{j}}^{2}}\beta_{jk}^{2}\\
&+\mathrm{log}(\alpha)\gamma_{jk}+\mathrm{log}(1-\alpha)(1-\gamma_{jk})\\
&+\mathrm{log}(\pi)\eta_{k}+\mathrm{log}(1-\pi)(1-\eta_{jk})\\
&+\mathrm{const}.
\end{aligned}
\end{eqnarray}
When $\eta_{k}=\gamma_{jk}=1\Leftrightarrow\eta_{k}\gamma_{jk}=1$, using Equation (34), we have
\begin{eqnarray}
  \begin{aligned}
  &\mathrm{log}\ q(\beta_{jk}|\eta_{k}=1,\gamma_{jk}=1)\\
  =&\left(-\frac{1}{2\sigma_{e_{j}}^{2}}\mathbf{x}_{jk}^{T}\mathbf{x}_{jk}-\frac{1}{2\sigma_{\beta_{j}}^{2}}\right)\beta_{jk}^{2}\\
  &+\left(\frac{\mathbf{x}_{jk}^{T}(\mathbf{y}_{j}-\mathbf{Z}_{j}\bm{\omega}_{j})-\sum_{k'\neq k}^{K}\mathbb{E}_{k'\neq k}[\eta_{k'}\gamma_{jk'}\beta_{jk'}]\mathbf{x}_{jk}^{T}\mathbf{x}_{jk'}}{\sigma_{e_{j}}^{2}}\right)\beta_{jk}\\
  &+\mathrm{const},
  \end{aligned}
\end{eqnarray}
from which we can see that the conditional posterior $q(\beta_{jk}|\eta_{k}=1,\gamma_{jk}=1)\sim \mathcal{N}(\mu_{jk},s_{jk}^{2})$, where
\begin{eqnarray}
  \begin{aligned}
  s_{jk}^{2}&=\frac{\sigma_{e_{j}}^{2}}{\mathbf{x}_{jk}^{T}\mathbf{x}_{jk}+\frac{\sigma_{e_{j}}^{2}}{\sigma_{\beta_{j}}^{2}}}\\
  \mu_{jk}&=\frac{\mathbf{x}_{jk}^{T}(\mathbf{y}_{j}-\mathbf{Z}_{j}\bm{\omega}_{j})-\sum_{k'\neq k}^{K}\mathbb{E}_{k'\neq k}[\eta_{k'}\gamma_{jk'}\beta_{jk'}]\mathbf{x}_{jk}^{T}\mathbf{x}_{jk'}}{\mathbf{x}_{jk}^{T}\mathbf{x}_{jk}+\frac{\sigma_{e_{j}}^{2}}{\sigma_{\beta_{j}}^{2}}}.
  \end{aligned}
\end{eqnarray}
For $\eta_{k}\gamma_{jk}=0$, we have
\begin{equation}
\mathrm{log}\ q(\beta_{jk}|\eta_{k}\gamma_{jk}=0)=-\frac{1}{2\sigma_{\beta_{j}}^{2}}\beta_{jk}^{2}+\mathrm{const},
\end{equation}
which implies that $q(\beta_{jk}|\eta_{k}\gamma_{jk}=0)\sim \mathcal{N}(0,\sigma_{\beta_{j}}^{2})$. Thus, the posterior is exactly the same as the prior if this variable is irrelevant in either one of the two levels ($\eta_{k}\gamma_{jk}=0$). Therefore we have
\begin{equation}
q(\eta,\gamma,\beta)=\prod_{k}^{K}\left(\pi_{k}^{\eta_{k}}(1-\pi_{k})^{1-\eta_{k}}\prod_{j}^{L}\left(\alpha_{jk}^{\gamma_{jk}}(1-\alpha_{jk})^{1-\gamma_{jk}}\mathcal{N}(\mu_{jk},s_{jk}^{2})^{\eta_{k}\gamma_{jk}}\mathcal{N}(0,\sigma^{2}_{\beta_{j}})^{1-\eta_{k}\gamma_{jk}}\right)\right),
\end{equation}
where we denote $\pi_k=q(\eta_k)$ and $\alpha_{jk}=q(\gamma_{jk})$.\\
Now we evaluate the second term of $\mathcal{L}(q)$ in Equation (32):

\begin{eqnarray}
  \begin{aligned}
  &-\mathbb{E}_q[\mathrm{log}q(\bm{\eta},\bm{\gamma},\bm{\beta})]\\
  =&-\mathbb{E}_{q}\left[\sum_{k}^{K}\left(\eta_{k}\mathrm{log}(\pi_{k})+(1-\eta_{k})\mathrm{log}(1-\pi_{k})\right)\right]\\
  &-\mathbb{E}_{q}\left[\sum_{k}^{K}\sum_{j}^{L}\eta_{k}\gamma_{jk}\mathrm{log}\mathcal{N}(\mu_{jk},s_{jk}^{2})+(1-\eta_{k}\gamma_{jk})\mathcal{N}(0,\sigma_{\beta_{j}}^{2})+\gamma_{jk}\mathrm{log}(\alpha_{jk})+(1-\gamma_{jk})\mathrm{log}(1-\alpha_{jk})\right]\\
  =&-\sum_{k}^{K}\mathbb{E}_{q}\left[\left(\eta_{k}\mathrm{log}(\pi_{k})+(1-\eta_{k})\mathrm{log}(1-\pi_{k})\right)\right]\\
  &-\sum_{k}^{K}\sum_{j}^{L}\mathbb{E}_{q}\left[\eta_{k}\gamma_{jk}\mathrm{log}\mathcal{N}(\mu_{jk},s_{jk}^{2})+(1-\eta_{k}\gamma_{jk})\mathcal{N}(0,\sigma_{\beta_{j}}^{2})+\gamma_{jk}\mathrm{log}(\alpha_{jk})+(1-\gamma_{jk})\mathrm{log}(1-\alpha_{jk})\right]\\
  =&-\sum_{k}^{K}\sum_{j}^{L}\mathbb{E}_{\eta_{k},\gamma_{jk}}\{\mathbb{E}_{\beta|\eta_{k}=1,\gamma_{jk}=1}[\mathrm{log}\mathcal{N}(\mu_{jk},s_{jk}^{2})]+\mathbb{E}_{\beta|\eta_{k}=1,\gamma_{jk}=0}[\mathrm{log}\mathcal{N}(0,\sigma_{\beta_{j}}^{2})]\\
  &+\mathbb{E}_{\beta|\eta_{k}=0,\gamma_{jk}=1}[\mathrm{log}\mathcal{N}(0,\sigma_{\beta_{j}}^{2})]+\mathbb{E}_{\beta|\eta_{k}=0,\gamma_{jk}=0}[\mathrm{log}\mathcal{N}(0,\sigma_{\beta_{j}}^{2})]\}\\
  &-\sum_{k}^{K}\sum_{j}^{L}\mathbb{E}_{q}[\gamma_{jk}\mathrm{log}(\alpha_{jk})+(1-\gamma_{jk})\mathrm{log}(1-\alpha_{jk})]-\sum_{k}^{K}\mathbb{E}_{q}[\eta_{k}\mathrm{log}(\pi_{k})+(1-\eta_{k})\mathrm{log}(1-\pi_{k})].\\
  \end{aligned}
\end{eqnarray}
Note that $-\mathbb{E}_{\beta|\eta_{k}=1,\gamma_{jk}=1}[\mathrm{log}\mathcal{N}(\mu_{jk},s_{jk}^{2})]$ is the entropy of Gaussian, so we have\\ $-\mathbb{E}_{\beta|\eta_{k}=1,\gamma_{jk}=1}[\mathrm{log}\mathcal{N}(\mu_{jk},s_{jk}^{2})]=\frac{1}{2}\mathrm{log}(s_{jk}^{2})+\frac{1}{2}(1+\mathrm{log}(2\pi))$, similarly, $-\mathbb{E}_{\beta|\eta_{k}=1,\gamma_{jk}=0}[\mathrm{log}\mathcal{N}(\mu_{jk},s_{jk}^{2})]=\frac{1}{2}\mathrm{log}(\sigma_{\beta_{j}}^{2})+\frac{1}{2}(1+\mathrm{log}(2\pi))$ and so on. Consequently, 
\begin{eqnarray}
  \begin{aligned}
  &-\mathbb{E}_q[\mathrm{log}q(\bm{\eta},\bm{\gamma},\bm{\beta})]\\
  =&\sum_{k}^{K}\sum_{j}^{L}\{[\frac{1}{2}\mathrm{log}(s_{jk}^{2})+\frac{1}{2}(1+\mathrm{log}(2\pi))]\pi_{k}\alpha_{jk}+[\frac{1}{2}\mathrm{log}(\sigma_{\beta_{j}}^{2})+\frac{1}{2}(1+\mathrm{log}(2\pi))](1-\pi_{k}\alpha_{jk})\\
  &-\alpha_{jk}\mathrm{log}(\alpha_{jk})-(1-\alpha_{jk})\mathrm{log}(1-\alpha_{jk})\}-\sum_{k}^{K}\{\pi_{k}\mathrm{log}(\pi_{k})+(1-\pi_{k})\mathrm{log}(1-\pi_{k})\}\\
  =&\sum_{k}^{K}\sum_{j}^{L}\frac{1}{2}\pi_{k}\alpha_{jk}(\mathrm{log}s_{jk}^{2}-\mathrm{log}\sigma_{\beta_{j}}^{2})+\frac{K}{2}\sum_{j}^{L}\mathrm{log}(\sigma_{\beta_{j}}^{2})+\frac{p}{2}+\frac{p}{2}\mathrm{log}(2\pi)\\
  &-\sum_{k}^{K}\sum_{j}^{L}[\alpha_{jk}\mathrm{log}(\alpha_{jk})+(1-\alpha_{jk})\mathrm{log}(1-\alpha_{jk})]-\sum_{k}^{K}[\pi_{k}\mathrm{log}(\pi_{k})+(1-\pi_{k})\mathrm{log}(1-\pi_{k})].
  \end{aligned}
\end{eqnarray}
Combine Equation (42) and Equation (35), we can find the lower bound:
\begin{eqnarray}
  \begin{aligned}
  &\mathbb{E}_q[\mathrm{log}\mathrm{Pr}(\mathbf{y}_{j},\bm{\eta},\bm{\gamma},\bm{\beta}|\mathbf{X};\theta)]-\mathbb{E}_q[\mathrm{log}q(\bm{\eta},\bm{\gamma},\bm{\beta})] \\
  =&\sum_{j=1}^{L}\Bigg\{-\frac{N_{j}}{2}\mathrm{log}(2\pi\sigma_{e_{j}}^{2})-\frac{\mathbf{y}_{j}^{T}\mathbf{y}_{j}}{2\sigma_{e_{j}}^{2}}-\frac{(\mathbf{Z}_{j}\bm{\omega}_{j})^{T}(\mathbf{Z}_{j}\bm{\omega}_{j})}{2\sigma_{e_{j}}^{2}}\\
  &+\frac{\sum_{k}^{K}\mathbb{E}_q\left[\eta_{k}\gamma_{jk}\beta_{jk}\right]\mathbf{x}_{jk}^{T}\mathbf{y}_{j}}{\sigma_{e_{j}}^{2}}+\frac{\mathbf{y}_{j}^{T}(\mathbf{Z}_{j}\bm{\omega}_{j})}{\sigma_{e_{j}}^{2}}-\frac{\sum_{k}^{K}\mathbb{E}_q\left[\eta_{k}\gamma_{jk}\beta_{jk}\right]\mathbf{x}_{jk}^{T}(\mathbf{Z}_{j}\bm{\omega}_{j})}{\sigma_{e_{j}}^{2}}\\
  &-\frac{1}{2\sigma_{e_{j}}^{2}}\sum_{k}^{K}\left(\mathbb{E}_q\left[\left(\eta_{k}\gamma_{jk}\beta_{jk}\right)^{2}\right]\mathbf{x}_{jk}^{T}\mathbf{x}_{jk}\right)\\
  &-\frac{1}{2\sigma_{e_{j}}^{2}}\left(\sum_{k}^{K}\sum_{k^{\prime}\neq k}^{K}\mathbb{E}_q\left[\eta_{k^{\prime}}\gamma_{jk^{\prime}}\beta_{j k^{\prime}}\right]\mathbb{E}_q\left[\eta_{k}\gamma_{jk}\beta_{jk}\right]\mathbf{x}_{j k^{\prime}}^{T}\mathbf{x}_{jk}\right)\Bigg\}  \\
  &-\frac{K}{2}\sum_{j}^{L}\mathrm{log}(2\pi\sigma_{\beta_{j}}^{2})-\sum_{k=1}^{K}\frac{\sum_{j=1}^{L}\mathbb{E}_q\left[\beta_{jk}^{2}\right]}{2\sigma_{\beta_{j}}^{2}}  \\
  &+\mathrm{log}(\alpha)\sum_{k=1}^{K}\sum_{j=1}^{L}\mathbb{E}_q\left[\gamma_{jk}\right]+\mathrm{log}(1-\alpha)\sum_{k=1}^{K}\sum_{j=1}^{L}\mathbb{E}_q\left[1-\gamma_{jk}\right]  \\
  &+\mathrm{log}(\pi)\sum_{k=1}^{K}\mathbb{E}_q\left[\eta_{k}\right]+\mathrm{log}(1-\pi)\sum_{k=1}^{K}\mathbb{E}_q\left[1-\eta_{k}\right]\\
  &+\sum_{k}^{K}\sum_{j}^{L}\frac{1}{2}\pi_{k}\alpha_{jk}(\mathrm{log}s_{jk}^{2}-\mathrm{log}\sigma_{\beta_{j}}^{2})+\frac{K}{2}\sum_{j}^{L}\mathrm{log}(\sigma_{\beta_{j}}^{2})+\frac{p}{2}+\frac{p}{2}\mathrm{log}(2\pi)\\
  &-\sum_{k}^{K}\sum_{j}^{L}[\alpha_{jk}\mathrm{log}(\alpha_{jk})]-\sum_{k}^{K}\sum_{j}^{L}[(1-\alpha_{jk})\mathrm{log}(1-\alpha_{jk})]\\
  &-\sum_{k}^{K}[\pi_{k}\mathrm{log}(\pi_{k})]-\sum_{k}^{K}[(1-\pi_{k})\mathrm{log}(1-\pi_{k})].
  \end{aligned}
\end{eqnarray}
Again we can show with the same technique in Section 1.1 that that $\mathbb{E}_q\left[\eta_{k}\gamma_{jk}\beta_{jk}\right]=\pi_{k}\alpha_{jk}\mu_{jk}$, $\mathbb{E}_q\left[(\eta_{k}\gamma_{jk}\beta_{jk})^{2}\right]=\pi_{k}\alpha_{jk}(s_{jk}^{2}+\mu_{jk}^{2})$, $\mathbb{E}_q\left[\beta_{jk}^{2}\right]=\pi_{k}\alpha_{jk}(s_{jk}^{2}+\mu_{jk}^{2})+(1-\pi_{k}\alpha_{jk})\sigma_{\beta_{j}}^{2}$, $\mathbb{E}_q\left[\eta_{k}\right]=\pi_{k}$, $\mathbb{E}_q\left[\gamma_{jk}\right]=\alpha_{jk}$. We plug in the expectations, then Equation (43) becomes\\
\begin{eqnarray}
  \begin{aligned}
  &\mathbb{E}_q[\mathrm{log}\mathrm{Pr}(\mathbf{y},\bm{\eta},\bm{\gamma},\bm{\beta}|\mathbf{X};\theta)]-\mathbb{E}_q[\mathrm{log}q(\bm{\eta},\bm{\gamma},\bm{\beta})] \\
  =&\sum_{j=1}^{L}\Bigg\{-\frac{N_{j}}{2}\mathrm{log}(2\pi\sigma_{e_{j}}^{2})-\frac{||\mathbf{y}_{j}-\mathbf{Z}_{j}\bm{\omega}_{j}-\sum_{k}^{K}\pi_{k}\alpha_{jk}\mu_{jk}\mathbf{x}_{jk}||^{2}}{2\sigma_{e_{j}}^{2}} \\
  &-\frac{1}{2\sigma_{e_{j}}^{2}}\sum_{k}^{K}\underbrace{[\pi_{k}\alpha_{jk}(s_{jk}^{2}+\mu_{jk}^{2})-(\pi_{k}\alpha_{jk}\mu_{jk})^{2}]}_{\text{Var}[\eta_{k}\gamma_{jk}\beta_{jk}]}\mathbf{x}_{jk}^{T}\mathbf{x}_{jk}\Bigg\}\\
  &-\frac{K}{2}\sum_{j=1}^{L}\mathrm{log}(2\pi\sigma_{\beta_{j}}^{2})-\frac{1}{2\sigma_{\beta_{j}}^{2}}\sum_{k=1}^{K}\sum_{j=1}^{L}[\pi_{k}\alpha_{jk}(s_{jk}^{2}+\mu_{jk}^{2})+(1-\pi_{k}\alpha_{jk})\sigma_{\beta_{j}}^{2}]\\
  &+\sum_{k=1}^{K}\sum_{j=1}^{L}\alpha_{jk}\mathrm{log}(\frac{\alpha}{\alpha_{jk}})+\sum_{k=1}^{K}\sum_{j=1}^{L}(1-\alpha_{jk})\mathrm{log}(\frac{1-\alpha}{1-\alpha_{jk}})\\
  &+\sum_{k=1}^{K}\pi_{k}\mathrm{log}(\frac{\pi}{\pi_{k}})+\sum_{k=1}^{K}(1-\pi_{k})\mathrm{log}(\frac{1-\pi}{1-\pi_{k}})\\
  &+\sum_{k}^{K}\sum_{j}^{L}\frac{1}{2}\pi_{k}\alpha_{jk}\mathrm{log}(\frac{s_{jk}^{2}}{\sigma_{\beta_{j}}^{2}})+\frac{K}{2}\sum_{j}^{L}\mathrm{log}(\sigma_{\beta_{j}}^{2})+\frac{p}{2}+\frac{p}{2}\mathrm{log}(2\pi).
  \end{aligned}
\end{eqnarray}
To get $\pi_{k}$ and $\alpha_{jk}$, we let $$\frac{\partial\mathbb{E}_q[\mathrm{log}\mathrm{Pr}(\mathbf{y}_{j},\bm{\eta},\bm{\gamma},\bm{\beta}|\mathbf{X};\theta)]-\mathbb{E}_q[\mathrm{log}q(\bm{\eta},\bm{\gamma},\bm{\beta})]}{\partial\pi_{k}}=0,$$ $$\frac{\partial\mathbb{E}_q[\mathrm{log}\mathrm{Pr}(\mathbf{y}_{j},\bm{\eta},\bm{\gamma},\bm{\beta}|\mathbf{X};\theta)]-\mathbb{E}_q[\mathrm{log}q(\bm{\eta},\bm{\gamma},\bm{\beta})]}{\partial\alpha_{jk}}=0,$$ 
··which gives us
\begin{eqnarray}
  \begin{aligned}
  &\pi_{k}=\frac{1}{1+\mathrm{exp}(-u_{k})},\\
  \mathrm{where}\ u_{k}=&\mathrm{log}\frac{\pi}{1-\pi}+\frac{1}{2}\sum_{j}^{L}\alpha_{jk}\left(\mathrm{log}\frac{s_{jk}^{2}}{\sigma_{\beta_{j}}^{2}}+\frac{\mu_{jk}^{2}}{s_{jk}^{2}}\right);\\
  \mathrm{and}\ &\alpha_{jk}=\frac{1}{1+\mathrm{exp}(-v_{jk})},\\
  \mathrm{where}\  v_{jk}=&\mathrm{log}\frac{\alpha}{1-\alpha}+\frac{1}{2}\pi_{k}\left(\mathrm{log}\frac{s_{jk}^{2}}{\sigma_{\beta_{j}}^{2}}+\frac{\mu_{jk}^{2}}{s_{jk}^{2}}\right).
  \end{aligned}
\end{eqnarray}
The derivation is as follow:
\begin{eqnarray}
  \begin{aligned}
   u_{k}=&\mathrm{log}\frac{\pi}{1-\pi}+\frac{1}{2}\sum_{j}^{L}\alpha_{jk}\mathrm{log}\frac{s_{jk}^{2}}{\sigma_{\beta_{j}}^{2}}\\
   &+\sum_{j}^{L}\frac{1}{\sigma_{e_{j}}^{2}}\left\{\alpha_{jk}\mu_{jk}\mathbf{x}_{jk}^{T}\mathbf{y}_{j}-\alpha_{jk}\mu_{jk}\mathbf{x}_{jk}^{T}(\mathbf{Z}_{j}\bm{\omega}_{j})-\sum_{k^{\prime}\neq k}^{K}\pi_{k^{\prime}}\alpha_{jk^{\prime}}\mu_{jk^{\prime}}\alpha_{jk}\mu_{jk}\mathbf{x}_{jk^{\prime}}^{T}\mathbf{x}_{jk}\right\}\\
   &-\frac{1}{2\sigma_{e_{j}}^{2}}\sum_{j}^{L}\alpha_{jk}(s_{jk}^{2}+\mu_{jk}^{2})\mathbf{x}_{jk}^{T}\mathbf{x}_{jk}-\frac{1}{2\sigma_{\beta_{j}}}\sum_{j}^{L}\alpha_{jk}(s_{jk}^{2}+\mu_{jk}^{2})+\frac{1}{2}\sum_{j}^{L}\alpha_{jk}\\
   =&\mathrm{log}\frac{\pi}{1-\pi}+\frac{1}{2}\sum_{j}^{L}\alpha_{jk}\mathrm{log}\frac{s_{jk}^{2}}{\sigma_{\beta_{j}}^{2}}\\
   &+\sum_{j}^{K}\alpha_{jk}\mu_{jk}\underbrace{\left(\frac{\mathbf{x}_{jk}^{T}(\mathbf{y}_{j}-\mathbf{Z}_{j}\bm{\omega}_{j})-\sum_{k^{\prime}\neq k}^{K}\pi_{k^{\prime}}\alpha_{jk^{\prime}}\mu_{jk^{\prime}}\mathbf{x}_{jk^{\prime}}^{T}\mathbf{x}_{jk}}{\sigma_{e_{j}}^{2}}\right)}_{\mu_{jk}/s_{jk}^{2}}\\
   &-\frac{1}{2}\sum_{j}^{L}\alpha_{jk}(s_{jk}^{2}+\mu_{jk}^{2})\underbrace{\left(\frac{\mathbf{x}_{jk}^{T}\mathbf{x}_{jk}}{\sigma_{e_{j}}^{2}}+\frac{1}{\sigma_{\beta_{j}}^{2}}\right)}_{1/s_{jk}^{2}}+\frac{1}{2}\sum_{j}^{L}\alpha_{jk}\\
   =&\mathrm{log}\frac{\pi}{1-\pi}+\frac{1}{2}\sum_{j}^{L}\alpha_{jk}\mathrm{log}\frac{s_{jk}^{2}}{\sigma_{\beta_{j}}^{2}}+\sum_{j}^{L}\frac{\alpha_{jk}\mu_{jk}^{2}}{s_{jk}^{2}}-\sum_{j}^{L}\frac{\alpha_{jk}\mu_{jk}^{2}}{2s_{jk}^{2}}\\
   =&\mathrm{log}\frac{\pi}{1-\pi}+\frac{1}{2}\sum_{j}^{L}\alpha_{jk}\left(\mathrm{log}\frac{s_{jk}^{2}}{\sigma_{\beta_{j}}^{2}}+\frac{\mu_{jk}^{2}}{s_{jk}^{2}}\right)，\\
  \end{aligned}
\end{eqnarray}
where we have used Equation (38). Similarly, we can derive $v_{jk}$.
\subsubsection{M-step}
At M-step, we update the parameters \{$\sigma_{e_{j}}^{2}, \sigma_{\beta_{j}}^{2}, \pi, \alpha, \bm{\omega}_{j}$\}. First we consider $\sigma_{e_{j}}^{2}$, by setting
$\frac{\partial\mathbb{E}_q[\mathrm{log}\mathrm{Pr}(\mathbf{y},\bm{\eta},\bm{\gamma},\bm{\beta}|\mathbf{X},\mathbf{Z};\theta)]}{\partial\sigma_{e_{j}}^{2}}=0,$
we have
\begin{equation}
  \begin{aligned}
  \sigma_{e_{j}}^{2}=&\frac{||\mathbf{y}_{j}-\mathbf{Z}_{j}\bm{\omega}_{j}-\sum_{k}^{K}\pi_{k}\alpha_{jk}\mu_{jk}\mathbf{x}_{jk}||^{2}}{N_{j}}\\
  &+\frac{\sum_{k}^{K}[\pi_{k}\alpha_{jk}(s_{jk}^{2}+\mu_{jk}^{2})-(\pi_{k}\alpha_{jk}\mu_{jk})^{2}]\mathbf{x}_{jk}^{T}\mathbf{x}_{jk}}{N_{j}}.
  \end{aligned}
\end{equation}
For $\sigma_{\beta_{j}}^{2}$, set $\frac{\partial \mathcal{L}(q)}{\partial\sigma_{\beta_{j}}^{2}}=0$, we have
\begin{equation}
\sigma_{\beta_{j}}^{2}=\frac{\sum_{k}^{K}\pi_{k}\alpha_{jk}(s_{jk}^{2}+\mu_{jk}^{2})}{\sum_{k}^{K}\pi_{k}\alpha_{jk}}.
\end{equation}
Accordingly,
\begin{equation}
\alpha=\frac{1}{p}\sum_{k}^{K}\sum_{j}^{L}\alpha_{jk},
\end{equation}
\begin{equation}
\pi=\frac{1}{K}\sum_{k}^{K}\pi_{k}.
\end{equation}
\begin{equation}
\bm{\omega}_{j}=(\mathbf{Z}_{j}^{T}\mathbf{Z}_{j})^{-1}\mathbf{Z}_{j}^{T}(\mathbf{y}_{j}-\sum_{k}^{K}\pi_{k}\alpha_{jk}\mu_{jk}\mathbf{x}_{jk}).
\end{equation}

\end{document}